\documentclass{elsart}

\usepackage{graphicx}
\usepackage{amsmath}
\usepackage{amssymb}

\journal{Annals of Physics}

\begin{document}

\begin{frontmatter}

\title{Spontaneous symmetry breaking and response functions}
\author{A. Beraudo}, 
\author{A. De Pace}, 
\author{M. Martini} and
\author{A. Molinari}
\address{Dipartimento di Fisica Teorica dell'Universit\`a di Torino and \\ 
  Istituto Nazionale di Fisica Nucleare, Sezione di Torino, \\ 
  via P.Giuria 1, I-10125 Torino, Italy}

\begin{abstract}
We study the quantum phase transition occurring in an infinite homogeneous
system of spin $1/2$ fermions in a non-relativistic context.
As an example we consider neutrons interacting through a simple spin-spin
Heisenberg force. The two critical values of the coupling strength ---
signaling the onset into the system of a finite magnetization and of the total
magnetization, respectively --- are found and their dependence upon the range
of the interaction is explored.
The spin response function of the system in the region where the
spin-rotational symmetry is spontaneously broken is also studied.
For a ferromagnetic interaction the spin response along the direction of the
spontaneous magnetization occurs in the particle-hole continuum and displays,
for not too large momentum transfers, two distinct peaks.
The response along the direction orthogonal to the spontaneous magnetization
displays instead, beyond a softened and depleted particle-hole continuum, a
collective mode to be identified with a Goldstone boson of type II.
Notably, the random phase approximation on a Hartree-Fock basis accounts for
it, in particular for its quadratic --- close to the origin --- dispersion
relation. It is shown that the Goldstone boson contributes to the saturation of
the energy-weighted sum rule for $\approx25\%$ when the system becomes fully
magnetized (that is in correspondence of the upper critical value of the
interaction strength) and continues to grow as the interaction strength 
increases.
\end{abstract}


\end{frontmatter}

\section{Introduction}

As emphasized by Iachello \cite{Iac03}, atomic nuclei offer an exciting ground
to explore phase transitions occurring at different temperatures $T$.
Indeed, for these systems at $T=0$ a phase transition is signaled by the
change they undergo from spherical to ellipsoidal shapes; at $T\approx20$~MeV
evidence exists for a liquid-gas transition and, finally, at $T\approx200$~MeV
the deconfinement of the hadronic (nuclear) matter is conjectured to take
place. 

As it is well-known, at $T=0$ the phase transitions are referred to as {\em
quantum phase transitions}. In fact, they occur in correspondence to special
(critical) values of a coupling constant, viewed as a continuous variable,
entering into the nuclear Hamiltonian: thus, they are driven, rather than by
thermal fluctuations, as it is the case at finite $T$, by quantum fluctuations.

In connection with quantum fluctuations, besides atomic nuclei, also neutron
stars are systems worth to be explored. Indeed, the strong magnetic
field present in these systems might be generated by a strong ferromagnetic
core in the stellar interior. Although a lot of effort has been devoted to this
issue (see Refs.~\cite{Sto03,Isa04,Vid02,Fan01,Sar03} for some recent work), no
firm conclusions have been reached at present.

In this paper we do not attempt a realistic calculation of the spin
polarization of a neutron star, but rather we address general aspects of a
quantum phase transition in an infinite homogeneous system of interacting 
neutrons, viewing the strength $V_1$ of a spin-spin ferromagnetic
neutron-neutron force (admittedly schematic with no pretense of being
realistic) as a control parameter. 
In Ref.~\cite{Ber04} we have indeed found that a critical value of the 
control parameter, namely $V_{1c}^{\textrm{lower}}$, exists such that, for
$V_1\ge V_{1c}^{\textrm{lower}}$, an incipient ferromagnetism sets in into the
system.
We have computed $V_{1c}^{\textrm{lower}}$, following two different paths:
\begin{itemize}
\item Random Phase Approximation (RPA) on a Hartree-Fock (HF) basis (RPA-HF),
\item anomalous single-particle propagator formalism, 
\end{itemize}
the latter only in the case of a zero-range force, finding that they lead to
identical results.

As it is well-known, the onset of the ferromagnetic phase into an infinite,
homogeneous system is signaled, on the one hand, by the divergence of the
system spin response function at vanishing frequency in the limit $(\omega=
0,\vec{q}\to 0)$, corresponding, in coordinate space, to a very large spin-spin
correlation length at the critical point. On the other hand, in the broken
phase the order parameter $<\widehat{M}>$ ($\widehat{M}$ being the system's
magnetization operator) acquires a finite expectation value: in
Ref.~\cite{Ber04} we obtained the latter through the Stoner equation
\cite{Hua98}. Equivalently one says that a vanishing (or a finite) value of 
$<\widehat{M}>$ corresponds to a symmetric (or to a broken) vacuum,
respectively. 

A second critical value $V_{1c}^{\textrm{upper}}$ of the control parameter,
associated to the system's full magnetization or to a completely broken vacuum,
exists and was found in Ref.~\cite{Ber04} in the anomalous propagator framework
only for the case of a zero-range interaction. 
In the present work we first explore the impact of the finite range of the
force, embedded in a parameter $\lambda$, on $V_{1c}^{\textrm{upper}}$.
This task was already carried out in Ref.~\cite{Ber04} for
$V_{1c}^{\textrm{lower}}$, which was computed as a function of $\lambda/k_F$,
$k_F$ being the Fermi momentum of the system, in the RPA-HF framework.
Here we employ the anomalous propagator formalism which allows not only to test
the previous RPA-HF results for $V_{1c}^{\textrm{lower}}(\lambda/k_F)$, but
also to get the range dependence of the upper critical value.

A second issue addressed in this work relates to the nature of the collective
modes of the system \textit{in the broken vacuum}. In this connection one
should distinguish between collective modes in the (longitudinal) $z$-direction
(assumed to coincide with the one along which the system's spontaneous
magnetization occurs) and in the direction orthogonal to $z$ (transverse). 
In the latter case the collective excitations truly represents the Goldstone
modes of the system: these inevitably occur in presence of a spontaneous
breaking of a continuous symmetry (in the present case the spin-rotational
one). We shall later address the issue of the number of these modes: here it
suffices to say that in the present case just one Goldstone mode exists.

We compute the collective modes of the system in the RPA-HF framework,
using again, for the sake of simplicity, a zero-range ferromagnetic force and
determining as a preliminary the region in the $(\omega,q)$ plane where the
response of the system occurs: this turns out to be different for the modes
developing in the $z$-direction and for those expanding in the plane orthogonal
to $z$.

In the former the region results from the interplay of four parabolas,
characterized by the Fermi momenta $k_F^+$ and $k_F^-$, for spin up and spin
down particles, respectively, which identify the broken vacuum. The values of
$k_F^+$ and $k_F^-$ are set by the strength of the interaction $V_1$.  
The four parabolas eventually coalesce into two when the system becomes fully 
magnetized. In these kinematical domains the response function to a
longitudinal probe displays a collective behavior, however embedded in the
particle-hole continuum and hence damped. 
Specifically, when the system is partially magnetized, for not too large values
of $q$ --- namely where the Pauli principle is active --- the response displays
two maxima, one of which is strongly softened and enhanced with respect to the
free case, whereas the other is pushed up close to the upper boundary of the
response, where it almost coincides with the peak already displayed by the free response: this situation is reminiscent of the splitting of the giant dipole
resonance in deformed nuclei \cite{Boh75}, although the physical
interpretation of the upper peak is different in the two cases (see later). 
For larger $q$ --- where the Pauli principle is no longer felt --- the two
maxima merge into one, still somewhat softened with respect to the free case,
if $q$, while large, is not too large.

Remarkably, for a fully magnetized system --- that is for a completely broken
vacuum --- no collective mode in the $z$-direction turns out to be possible: an
occurrence, however, strictly valid only for a zero-range interaction.

Turning to the transverse collective modes, when the vacuum is broken by a
ferromagnetic spin-spin interaction a region free of single-particle response
opens up in the corner of the $(\omega,q)$ plane: here is where the Goldstone
boson lives. 
Actually the latter, as we shall see, displays --- in the combined RPA and
HF framework and for not too large momenta --- a parabolic (rather than linear)
dispersion relation: hence, according to the general theory of the spontaneous
symmetry breaking in the non-relativistic regime \cite{Nie76}, it turns out to
be a Goldstone mode of second kind. 
It is remarkable that the RPA-HF theory accounts for its existence through a
parabolic dispersion relation, which, however, for larger momenta deviates from
the parabolic behavior. Indeed, it develops, just before entering into the
particle-hole continuum,  a maximum that entails the vanishing of the group
velocity of the propagating mode. 

Much information on the collective motion in a many-body system can be gained
already through the moments of the response function, referred to as {\em sum
rules}. In the final part of this work we shall consider the zeroth ($S_0$) and
the first ($S_1$) moments, commonly called the Coulomb and the energy-weighted
sum rules, respectively. Specifically, we shall address the issue of finding
out how their value is affected by the amount of symmetry breaking occurring in
the vacuum, i.~e. in the system's ground state.

The present paper is organized as follows: in Section~\ref{sec:crit} we deal
with the problem of determining the critical values of the control parameters
and their dependence upon the range of the interaction.
In Section~\ref{sec:zaxis} we explore the response of the system in the
presence of a spontaneous symmetry breaking. Specifically, we analyze the
system's response in the direction (referred to as longitudinal) along which
its spontaneous magnetization occurs.
In Section~\ref{sec:xaxis} we study the system's response orthogonal to
the direction of the spontaneous magnetization (referred to as transverse).
Here the Goldstone nature of the system collective excited state is thoroughly
discussed. 
Finally, in Section~\ref{sec:moments}, the same issue is addressed in the
context of the moments of the system's response functions (the so-called {\em
sum rules}).
In the concluding section we summarize our work and shortly discuss the nature
of the quantum phase transitions addressed in our research. Also, we shortly 
discuss their significance for the physics of the atomic nuclei and of the
neutron stars. 

\section{The critical values of the control parameter}
\label{sec:crit}

We have recalled in the previous Section that unlike the symmetric vacuum, 
which is specified by just one 
Fermi momentum $k_F$,
the broken vacuum of our system is characterized by two Fermi momenta
$k_F^+$ and $k_F^-$: these fix the densities of the particles with spin up and
spin down, respectively. The equilibrium condition for our system at zero
temperature reads then: 
\begin{equation}
  \omega^{+}_{k_F^+}=\omega^{-}_{k_F^-},
\label{equil}
\end{equation}
where the single-particle energies 
\begin{subequations}
\label{eq:omegapm}
\begin{eqnarray}
  \omega^{\pm}_{\vec{k}} &=& \omega_{\vec{k}} + \Sigma_{\pm\pm}(k) \\
  \omega_{\vec{k}} &=& \frac{k^2}{2m}
\end{eqnarray}
\end{subequations}
embody the exact Hartree-Fock (HF) self-energy\footnote{We remind the reader
  that, although our interaction is of pure exchange character, an Hartree term
  arises in the self-energy owing to the broken vacuum (see
  Ref.~\protect\cite{Ber04}).} that, while diagonal, is no longer proportional
to the unit matrix. Indeed for the interaction (in momentum space)
\begin{equation}
\label{finite_V_k}
  V(k)=V_1\vec{\sigma}_{1} \cdot
  \vec{\sigma}_{2}\frac{\lambda^2}{k^2+\lambda^2} 
\end{equation}
the anomalous HF self-energy turns out to read
\begin{subequations}
\begin{eqnarray}
\label{sigma+}
  \Sigma_{++}(k_F^{+},k_F^{-},k)&=&\frac{V_1}{6\pi^2}
    \left[(k_F^{+})^3-(k_F^{-})^3\right] \nonumber\\
  && -\frac{V_1\lambda^3}{(2\pi)^2} \left[\frac{k_F^+}{\lambda}-
    \left(\arctan\frac{k_F^+-k}{\lambda}+
    \arctan\frac{k_F^++k}{\lambda}\right)\right. \nonumber\\
  && +\frac{1}{4}\left(\frac{(k_F^{+})^2}{\lambda k}-\frac{k}{\lambda}
    +\frac{\lambda}{k}\right)
    \ln\frac{1+(k_F^++k)^2/\lambda^2}{1+(k_F^+-k)^2/\lambda^2} \nonumber\\
  && +2\left(\frac{k_F^-}{\lambda}-\left(\arctan\frac{k_F^--k}{\lambda}
    +\arctan\frac{k_F^-+k}{\lambda}\right)\right. \nonumber\\
  && +\left.\left.\frac{1}{4}\left(\frac{(k_F^{-})^2}{\lambda k}-
    \frac{k}{\lambda}+\frac{\lambda}{k}\right)
    \ln\frac{1+(k_F^-+k)^2/\lambda^2}{1+(k_F^--k)^2/\lambda^2}\right)\right]
    \\\nonumber
\end{eqnarray}
for spin up neutrons and 
\begin{eqnarray}
\label{sigma-}
  \Sigma_{--}(k_F^{+},k_F^{-},k)&=&\Sigma_{++}(k_F^{-},k_F^{+},k)
\end{eqnarray}
\end{subequations}
for the spin down ones.\\
Then, inserting (\ref{sigma+}) and (\ref{sigma-}), evaluated at the appropriate
value of $k$, into Eq.~(\ref{equil}) one gets the equation
\begin{eqnarray}
  &&\frac{1}{2m}\left[(k_F^{+})^2-(k_F^{-})^2\right]=
    \frac{V_1}{3\pi^2}\left[(k_F^{+})^3-(k_F^{-})^3\right] \nonumber \\
  && \qquad +\frac{V_1\lambda^3}{(2\pi)^2}
    \left[\frac{k_F^--k_F^+}{\lambda}+\arctan\frac{2k_F^-}{\lambda}
    -\arctan\frac{2k_F^+}{\lambda}
    +4\arctan\frac{k_F^+-k_F^-}{\lambda}\right. \nonumber \\
  && \qquad +\frac{\lambda}{4k_F^+}
    \ln\left[1+\left(\frac{2k_F^+}{\lambda}\right)^2\right]
    -\frac{\lambda}{4k_F^-}
    \ln\left[1+\left(\frac{2k_F^-}{\lambda}\right)^2\right] \nonumber \\
  && \qquad +\left.\frac{(k_F^--k_F^+)[(k_F^++k_F^-)^2+\lambda^2]}
    {2\lambda k_F^+ k_F^-}\ln\frac{1+(k_F^++k_F^-)^2/\lambda^2}
    {1+(k_F^+-k_F^-)^2}\right],
\label{critico}
\end{eqnarray}
which, in the case of a zero-range force, reduces to the simple expression
\begin{equation}
  V_1=-\frac{\pi^2}{m}\frac{(k_F^{+})^2-(k_F^{-})^2}
    {(k_F^{+})^3-(k_F^{-})^3}.
\label{eq:zerorange}
\end{equation}
Writing in the above the Fermi momenta $k_F^+$ and $k_F^{-}$ in terms 
of the magnetization $M$ of the system according to
\begin{equation}
  k_F^{ \pm }=[3\pi^2\rho(1\pm M)]^{1/3}
\end{equation}
one obtains the Stoner equation for the critical value of the coupling $V_1$. 
In the above
\begin{equation}
  \rho=\frac{k_F^3}{3\pi^2}=\frac{(k_F^+)^3}{6\pi^2}+\frac{(k_F^-)^3}{6\pi^2}
\label{eq:density}
\end{equation}
is the density of the system and, of course, $0\le M\le 1$.

\begin{figure}
\begin{center}
\includegraphics[clip,height=5.7cm]{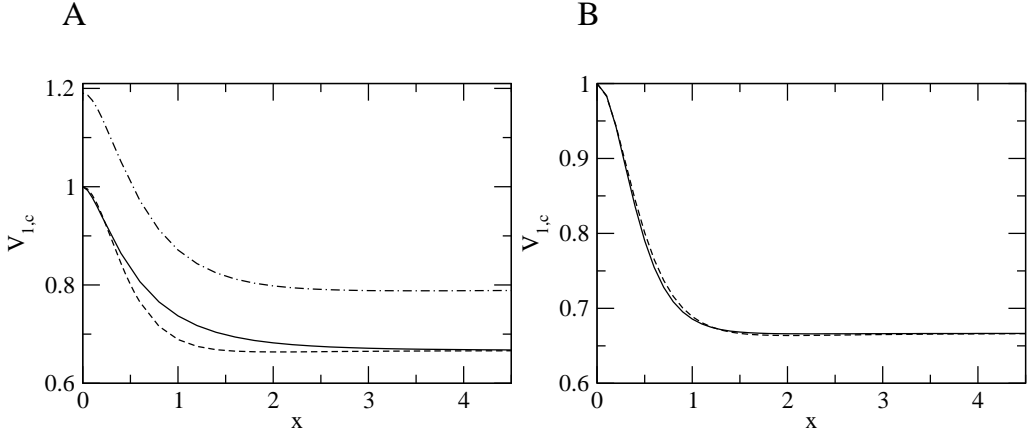}
\caption{\label{fig:comp_vcrit}In panel (A) the behavior of the critical 
  couplings $V_{1,c}^{\textrm{upper}}$ (dot-dashed) and
  $V_{1,c}^{\textrm{lower}}$ (solid) computed in the anomalous propagator
  formalism are shown versus $x=\lambda/k_F$; also displayed is
  $V_{1,c}^{\textrm{lower}}$ in RPA-HF (dashed).  
  In panel (B) one sees $V_{1,c}^{\textrm{lower}}$ in the effective mass 
  approximation both in the anomalous propagator framework (solid) and in
  RPA-HF (dashed). All curves are in units of $-\pi^2/mk_F$. } 
\end{center}
\end{figure}

Solving Eq.~(\ref{critico}) for $M=1$, which implies $k_F^{+}=2^{1/3}k_F$, 
and setting $x=\lambda/k_F$ one gets
\begin{eqnarray}
  &&V_{1,c}^{\textrm{upper}}=-\frac{\pi^2}{mk_F} \frac{1}{2^{1/3}}
    \left[\frac{2}{3}+\frac{x^2}{x^2+2^{2/3}} +\frac{x^2}{2~2^{2/3}}
    +\frac{x^2}{2^{2/3}}\frac{x^2}{x^2+2^{2/3}}\right. \nonumber \\
  && \ \left.-x^3\arctan\left(\frac{2^{1/3}}{x}\right) +\frac{x^3}{4}
    \arctan\left(\frac{2~2^{1/3}}{x}\right)
    -\frac{x^4}{2^3~2^{4/3}}\ln\left(1+4\frac{2^{2/3}}{x^2}\right)
    \right]^{-1}. \nonumber \\
\end{eqnarray}
This formula, though cumbersome, becomes transparent for a zero 
range interaction ($x\to\infty$), where it reads
\begin{equation}
  \lim_{x\to\infty}V_{1,c}^{\textrm{upper}}=-\frac{\pi^2}{mk_F}\frac{1}
    {2^{1/3}},
\end{equation}
a value coinciding with the one found in Ref.~\cite{Ber04}, while for an 
infinite range force ($x\to 0$), where it yields
\begin{equation}
  \lim_{x\to 0}V_{1,c}^{\textrm{upper}}=-\frac{\pi^2}{mk_F}
  \frac{1}{2^{1/3}}\frac{3}{2}.
\end{equation}
The behavior of the absolute value of $V_{1,c}^{\textrm{upper}}$ is shown 
in Fig.~\ref{fig:comp_vcrit} (panel A). Note that the actual strength of the
interaction (\ref{finite_V_k}) is $V_1 \lambda^2$: it would display a
monotonically increasing behavior with $\lambda$, showing that the larger
the range of the force, the smaller the resistance of the medium to become
fully magnetized. It is worth reminding that this behavior is shared by the
$O(3)$ model \cite{Car96}, which enjoys the same symmetry of ours and
represents a generalization of the Ising model \cite{Hua87}.
Also remarkable is that for ranges of the force smaller than about
$(2k_F)^{-1}$ the critical value of the control parameter becomes essentially
insensitive to $\lambda$.

In concluding this Section we quote the lower critical value of the coupling
$V_1$ as obtained through the Stoner Equation. For this purpose it merely
suffices to solve Eq.~(\ref{critico}) setting $<\widehat{M}>=0$. One gets
\begin{equation}
\label{eq:V1clower}
  V_{1,c}^{\textrm{lower}}=-\frac{\pi^2}{mk_F}\left[1+\frac{x^2}{2} 
    (1+\frac{3}{8}x^2) \ln(1+\frac{4}{x^2})-\frac{3}{4}x^2\right]^{-1},
\end{equation}
which for a zero-range interaction reduces to
\begin{equation}
  \lim_{x\to\infty}V_{1,c}^{\textrm{lower}}=
    -\frac{\pi^2}{mk_F}\cdot\frac{2}{3}, 
\end{equation}
a value coinciding with what found in \cite{Ber04} and for an infinite range
force yields 
\begin{equation}
  \lim_{x\to 0}V_{1,c}^{\textrm{lower}}=-\frac{\pi^2}{mk_F}.
\end{equation}
Formula (\ref{eq:V1clower}) is displayed in Fig.~\ref{fig:comp_vcrit} (panel A)
as a function of $x$. In the same figure it is also shown the result obtained
in Ref.~\cite{Ber04} through the solution of the RPA equation in the long wave
length, small frequency domain \cite{Alb82}. 
Notably the two curves differ at intermediate value of $x$. 
This outcome relates to the approximation made in Ref.~\cite{Ber04}, namely of
using the effective mass as an approximation to the self-consistent HF
solution. Indeed, if we do the same in the present anomalous propagator
framework, we find for the neutron self energy
\begin{equation}
  \Sigma_{++}^{\textrm{eff}}(k_F^{+},k_F^{-},k)= A_{++}+B_{++}k^2 ,
\end{equation}
with
\begin{subequations}
\begin{eqnarray}
  A_{++}&=&\frac{V_1}{6\pi^2}\left[(k_F^{+})^3-(k_F^{-})^3\right] \nonumber\\
  && -\frac{V_1\lambda^3}{(2\pi)^2}
    \left[\left(\frac{k_F^+}{\lambda}-\arctan\frac{k_F^+}{\lambda}\right)+
    2\left(\frac{k_F^-}{\lambda}-\arctan\frac{k_F^-}{\lambda}\right)\right] \\
  B_{++}&=&\frac{V_1\lambda^2}{6\pi^2}\left[\frac{(k_F^{+})^3}
    {((k_F^{+})^2+\lambda^2)^2}+2\frac{(k_F^{-})^3}
    {((k_F^{-})^2+\lambda^2)^2}\right]
\label{sigmameff}
\end{eqnarray}
\end{subequations}
and  
\begin{equation}
  \Sigma_{--}^{\textrm{eff}}(k_F^{+},k_F^{-},k)=
    \Sigma_{++}^{\textrm{eff}}(k_F^{-},k_F^{+},k). 
\end{equation}
It is gratifying to see in Fig.~\ref{fig:comp_vcrit} (panel B) that now, namely
in the effective mass approximation, the anomalous propagator framework and the
RPA-HF lead to identical results. 

\section{The system's longitudinal response}
\label{sec:zaxis}

Let us assume the system to undergo a spontaneous symmetry breaking 
acquiring a magnetization along the $z$-axis.
We wish to explore the system's response to a spin-dependent, but not
spin-flipping, external probe acting in the $z$ direction, namely described by
an operator 
\begin{equation}
\label{ext_probe}
  \widehat{O}=\sum_{r\beta \atop s\alpha}<r\beta|\sigma_z|s\alpha>
  \hat{a}^{\dag}_{r\beta}\hat{a}_{s\alpha}=
  \sum_{s\alpha}\hat{a}^{\dag}_{s\alpha}\hat{a}_{s\alpha} (-1)^{1/2-\alpha}.
\end{equation}
For sake of simplicity we confine ourselves to assume a ferromagnetic
($V_1<0$), spin-dependent, zero-range interaction among neutrons, namely 
\begin{equation}
\label{Vdelta}
  V(r)=V_1 \vec{\sigma}_{1} \cdot \vec{\sigma}_{2} \delta(r),
\end{equation}
clearly constant in momentum space (see Eq.(\ref{finite_V_k}) in 
$\lambda\to\infty$ limit).

We shall compute the response to the probe (\ref{ext_probe}) in the RPA-HF 
framework both in a normal and in a broken vacuum.
To this end it is first necessary to set up the longitudinal HF anomalous
polarization propagator in the broken vacuum $\Pi^{\textrm{HF},b}_{zz}$. 
This is easily achieved starting from the anomalous single-particle propagator 
(the label ``b'' stands for ``broken vacuum'')
\begin{equation}
\label{eq:G0matrix}
  G^{\textrm{HF},b}(\vec{k},k_0) = \left( 
    \begin{array}{cc}
      G^{\textrm{HF},b}_{++}(\vec{k},k_0) & 0 \\
      0 & G^{\textrm{HF},b}_{--}(\vec{k},k_0)
    \end{array} \right),
\end{equation}
where 
\begin{subequations}
\begin{eqnarray}
  G^{\mathrm{HF},\mathrm{b}}_{++}(\vec{k},k_0) &=& 
    \frac{\theta(k-k_F^+)}{k_0-\omega^{+}_{\vec{k}}+i\eta} + 
    \frac{\theta(k_F^+-k)}{k_0-\omega^{+}_{\vec{k}}-i\eta}
\label{eq:GHF++} 
\end{eqnarray}
is the propagator for a spin up neutron and
\begin{eqnarray}
  G^{\mathrm{HF},\mathrm{b}}_{--}(\vec{k},k_0) &=& 
    \frac{\theta(k-k_F^-)}{k_0-\omega^{-}_{\vec{k}}+i\eta} + 
    \frac{\theta(k_F^--k)}{k_0-\omega^{-}_{\vec{k}}-i\eta}
\label{eq:GHF--}
\end{eqnarray}
\end{subequations}
is the propagator for a spin down one
and 
\begin{equation}
  \omega^{\pm}_{\vec{k}} = \frac{k^2}{2m}-\frac{V_1 {k_F^{\mp}}^3}{2\pi^2} 
\end{equation}
are the single-particle energies (\ref{eq:omegapm}) for a zero-range force.
In the following we shall specify the spin indices for $G$ only, dropping all
the other ones, and we shall always assume the HF approximation for the
single-particle propagator. 

One gets then for the HF polarization propagator in the $z$-direction and in
the broken vacuum, setting $K\equiv(k_0,\vec{k})$ and
$Q\equiv(\omega,\vec{q})$, 
\begin{eqnarray}
\label{pizerobzz}
  \Pi^{\textrm{HF},b}_{zz}(Q) & = & - i
    \int\frac{{\d^4}K}{(2\pi)^4} [G_{++}(K)G_{++}(K+Q)+
    G_{--}(K)G_{--}(K+Q)] \nonumber \\
  &=& \Pi^{\textrm{HF}}_{++}(Q)+\Pi^{\textrm{HF}}_{--}(Q)
\end{eqnarray}
being
\begin{eqnarray}
\label{pizerobpiu}
  \Pi^{\textrm{HF}}_{++}(Q)=\Pi^{0}_{++}(Q) & = & 
    \int\frac{{\d}\vec{k}}{(2\pi)^3}
    \theta(|\vec{k}+\vec{q}|-k_F^+)\theta(k_F^+-k)\nonumber \\
  && \times   
    \left[\frac{1}{\omega+\omega_{\vec{k}}-\omega_{\vec{k}+\vec{q}}+i\eta}-
    \frac{1}{\omega+\omega_{\vec{k}+\vec{q}}-\omega_{\vec{k}}-i\eta}\right]
\end{eqnarray}
and
\begin{eqnarray}
\label{pizerobmeno}
  \Pi^{\textrm{HF}}_{--}(Q)=\Pi^{0}_{--}(Q) & = & 
    \int\frac{{\d}\vec{k}}{(2\pi)^3}
    \theta(|\vec{k}+\vec{q}|-k_F^-)\theta(k_F^--k)\nonumber \\
  && \times
    \left[\frac{1}{\omega+\omega_{\vec{k}}-\omega_{\vec{k}+\vec{q}}+i\eta}-
    \frac{1}{\omega+\omega_{\vec{k}+\vec{q}}-\omega_{\vec{k}}-i\eta}\right].
\end{eqnarray}
Note that the HF expressions for $\Pi_{++}$ and $\Pi_{--}$ are identical to the
free ones in the case of a zero-range interaction. Moreover both their real 
and imaginary part can easily be computed analytically: one clearly obtains 
the familiar results for a symmetric vacuum \cite{Fet71} with $k_F$ replaced by
$k_F^+$ and $k_F^-$, respectively. 

\begin{figure}
\begin{center}
\includegraphics[clip,height=7cm]{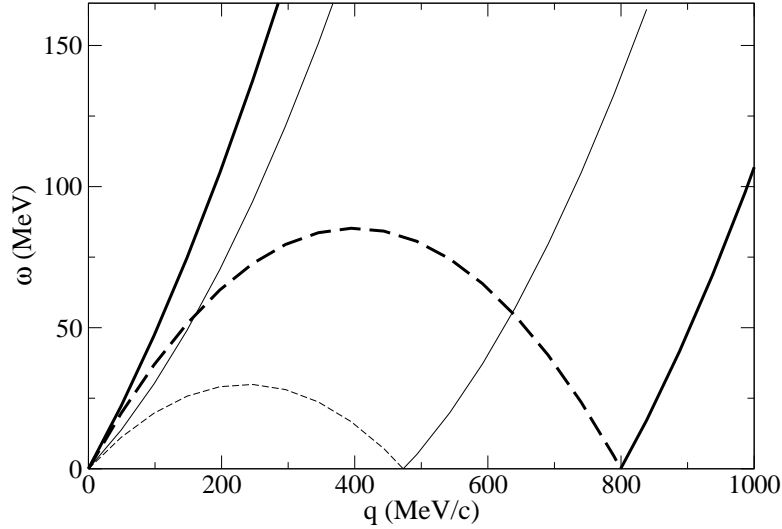}
\caption{\label{fig:resp_region} The response region in a broken vacuum.
  In the figure $k_F^+=400$ MeV/c and $k_F^-=237$ MeV/c. The heavy
  (light) line marks the domain where the spin up (down) neutrons respond 
  to an external probe. The dashed lines represent the boundaries of the Pauli
  blocked regions.} 
\end{center}
\end{figure} 
Eqs.~(\ref{pizerobzz}), (\ref{pizerobpiu}) and (\ref{pizerobmeno}) 
allow us to display in Fig.~\ref{fig:resp_region} the response region 
of the system, in the frequency $\omega$ - momentum $q$  plane, when the vacuum
is broken.
We see in the figure that the global response region is actually made up by two
response domains: one associated with $k_F^+$, where the particles with 
spin up respond to the external probe and the other associated with $k_F^-$,
where the particles with spin down respond to the external probe.

\begin{figure}
\begin{center}
\includegraphics[clip,height=6cm]{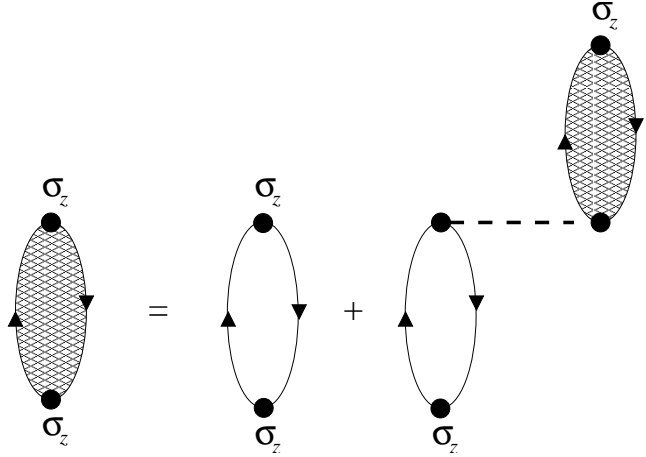}
\caption{\label{fig:piRPA}
  The Dyson equation for the RPA anomalous polarization propagator.} 
\end{center}
\end{figure}

Turning to the propagator $\Pi^{\textrm{RPA-HF},b}_{zz}$ in RPA-HF, it obeys
the equation
\begin{equation}
\label{trick}
  \widehat{\Pi}=\widehat{\Pi}^{\textrm{HF}} +
    \widehat{\Pi}^{\textrm{HF}}\widehat{V}\widehat{\Pi}, 
\end{equation} 
graphically displayed in Fig.~\ref{fig:piRPA}, where:
\begin{equation}
\label{eq:Pime}
  \widehat{\Pi} = \left( 
    \begin{array}{cc}
      \Pi_{++} & \Pi_{+-}\\
      \Pi_{-+} & \Pi_{--}
    \end{array}
    \right)\,\quad \widehat{\Pi}^{\textrm{HF}} = \left( 
    \begin{array}{cc}
      \Pi_{++}^{\textrm{HF}} & 0\\
      0 & \Pi_{--}^{\textrm{HF}}
    \end{array} 
    \right)\quad\textrm{and}\quad\widehat{V} = \left( 
    \begin{array}{cc}
      V_d & V_{od}\\
      V_{od} & V_d
    \end{array} 
    \right).
\end{equation}
The solution of Eq.~(\ref{trick}) turns out to read (see 
Appendix~\ref{app:rpazz})
\begin{equation}
\label{pibRPA}
  \Pi^{\textrm{RPA-HF},b}_{zz}=
    \frac{\Pi_{++}^{\textrm{HF}}+\Pi_{--}^{\textrm{HF}} + 
    2\Pi_{++}^{\textrm{HF}}\Pi_{--}^{\textrm{HF}} (V_{od}-V_d)}
    {1-V_d(\Pi_{++}^{\textrm{HF}} + \Pi_{--}^{\textrm{HF}}) +
    \Pi_{++}^{\textrm{HF}}\Pi_{--}^{\textrm{HF}}(V_d^2-V_{od}^2)}.
\end{equation}
In the above $V_d$ and $V_{od}$ correspond to the diagonal and off-diagonal 
particle-hole matrix elements of the interaction (\ref{Vdelta}) in spin space:
they are given in Appendix~\ref{app:rpazz}. Noteworthy is that
Eq.~(\ref{pibRPA}) is formally similar to what one gets in systems containing
two species of particles, e.g. nucleons and $\Delta$'s (see
Ref.~\cite{Alb89b}). 

\begin{figure}
\begin{center}
\includegraphics[clip,width=\textwidth]{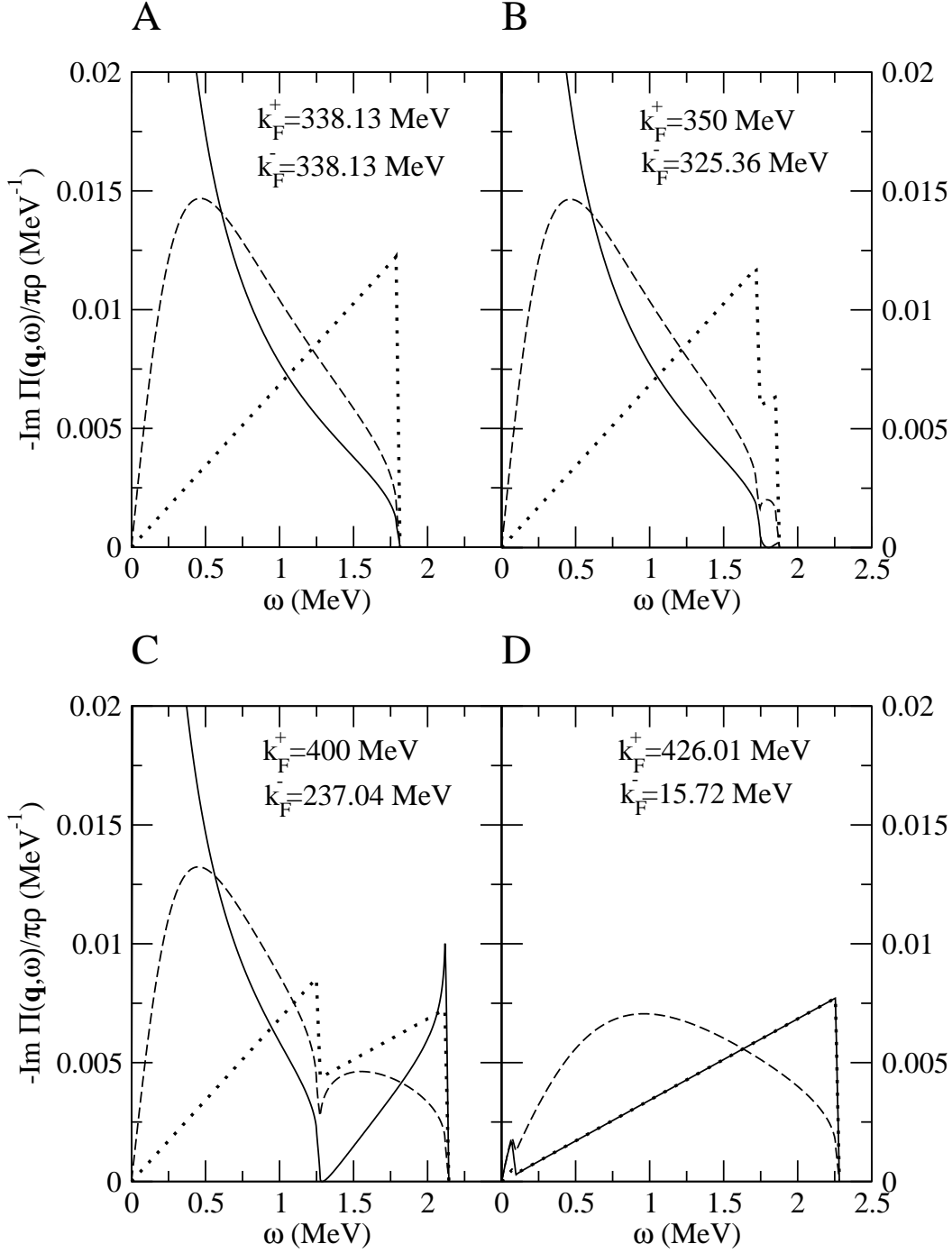}
\caption{\label{fig:Pizz_q005} The response of an infinite neutron's 
  system to a $z$-aligned probe for $q=5$ MeV/c. Panel (A) refers to the case 
  of a symmetric vacuum ($k_F=k_{F}^{+}=k_{F}^{-}$), panels (B) and (C) to 
  a partially aligned broken vacuum, panel (D) to a totally broken, fully 
  aligned vacuum. Dotted line: HF (free) response, dashed line: ring
  approximation, solid line: RPA-HF.} 
\end{center}
\end{figure} 
\begin{figure}
\begin{center}
\includegraphics[clip,width=\textwidth]{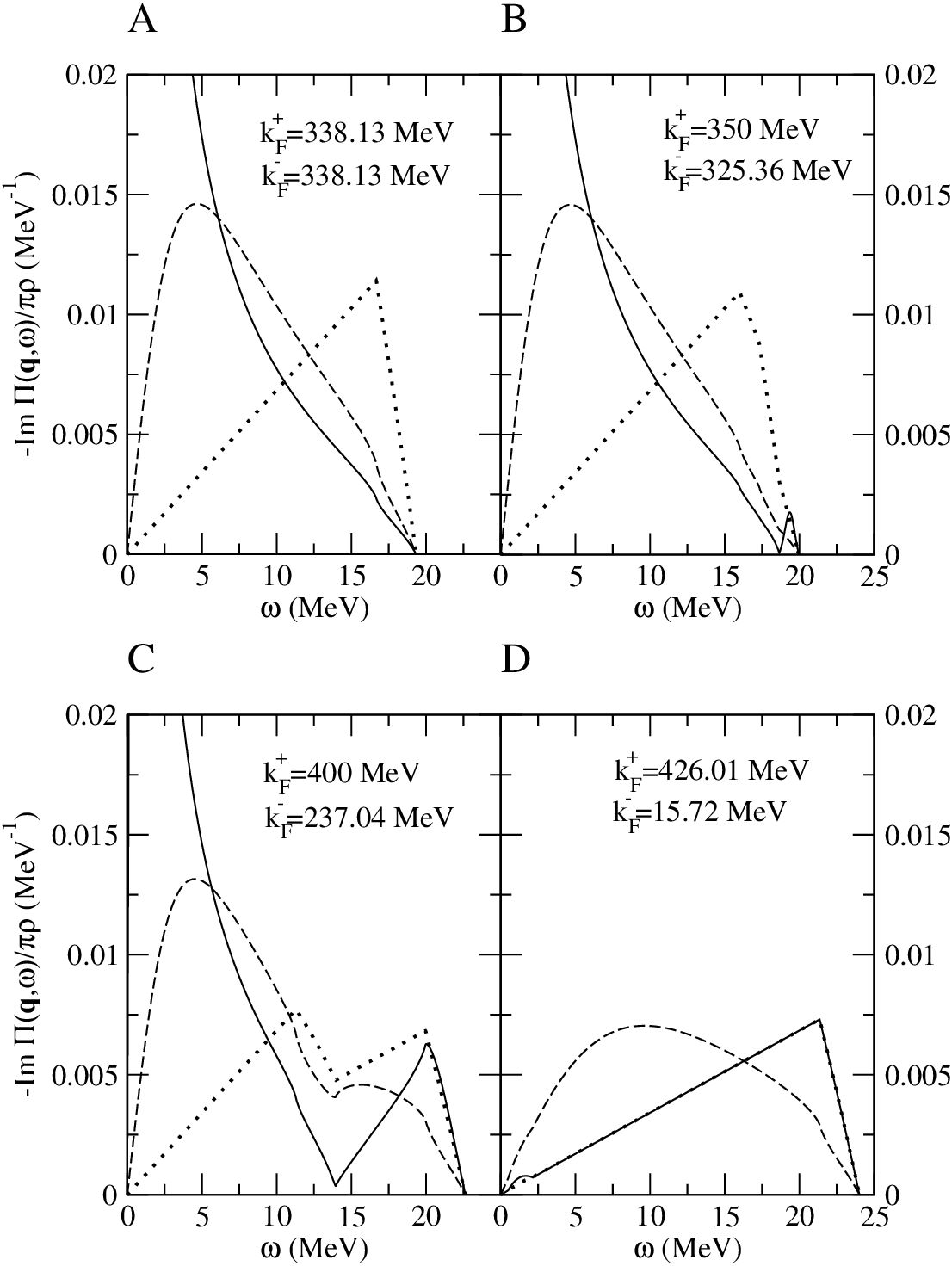}
\caption{\label{fig:Pizz_q050} The same as in Fig.~\protect\ref{fig:Pizz_q005}
  but for $q=50$ MeV/c. } 
\end{center}
\vskip 3cm
\end{figure} 
\begin{figure}
\begin{center}
\includegraphics[clip,width=\textwidth]{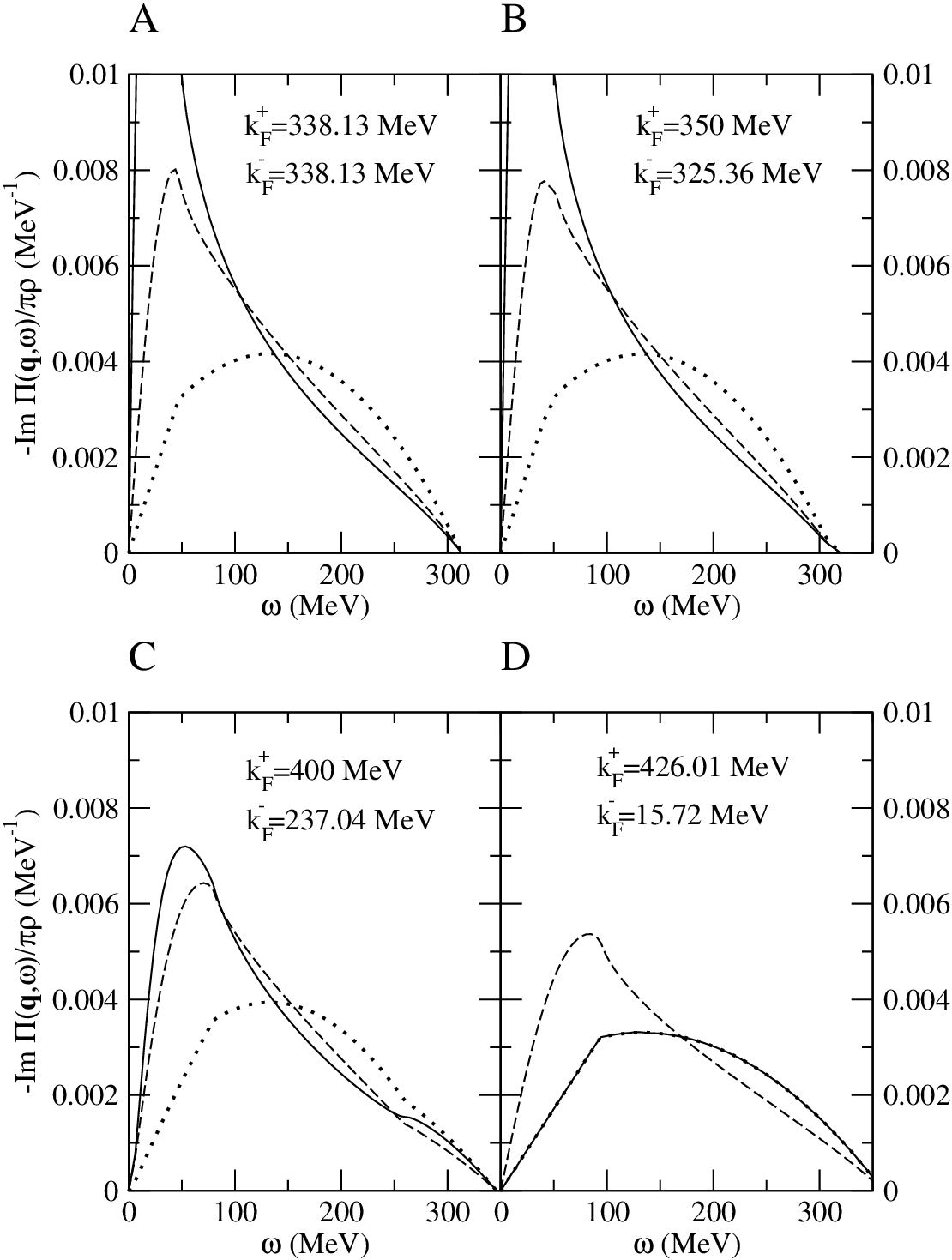}
\caption{\label{fig:Pizz_q500} As in Fig.~\protect\ref{fig:Pizz_q005}
  but for $q=500$ MeV/c. } 
\end{center}
\vskip 3cm
\end{figure} 
Equation (\ref{pibRPA}) entails a striking consequence, namely that 
for a fully broken vacuum (a fully magnetized system, for example in the 
positive $z$-direction) no RPA collective mode exists for a zero-range force.
Indeed, in this case, since $k_F^-=0$ then $\Pi^{0}_{--}=0$ and also, 
as shown in Appendix~\ref{app:rpazz}, $V_d$=0. 
The situation is clearly illustrated in Figs.~\ref{fig:Pizz_q005},
\ref{fig:Pizz_q050} and \ref{fig:Pizz_q500}, where the system's response 
along the $z$-axis is shown at $q=5$, $50$ and $500$~MeV/c, respectively.
Furthermore, in each figure, the evolution of the system's response with the 
amount of breaking of the vacuum is also displayed. Accordingly, in each figure
the responses associated to four pairs of values of $k_F^+$ and $k_F^-$ are
shown: since the density of the system is fixed, these are related by
Eq.~(\ref{eq:density}). 
We further observe that each choice of ($k_F^+$,$k_F^-$) corresponds to a value
of the strength of the interaction $V_1$ given by Eq.~(\ref{eq:zerorange}). 
In panel A of all figures the enhancement and softening of the response 
in the symmetric vacuum ($k_F^+=338.13$ MeV/c), due 
to the attractive ferromagnetic interaction, is clearly apparent.
As one moves towards an increasingly broken vacuum and for not too large
momenta one sees the appearance of a second peak in the response at high energy
until, for a totally broken vacuum, the collectivity completely disappears in
accord with the argument given above and the free response is recovered.

\begin{figure}
\begin{center}
\includegraphics[clip,height=7cm]{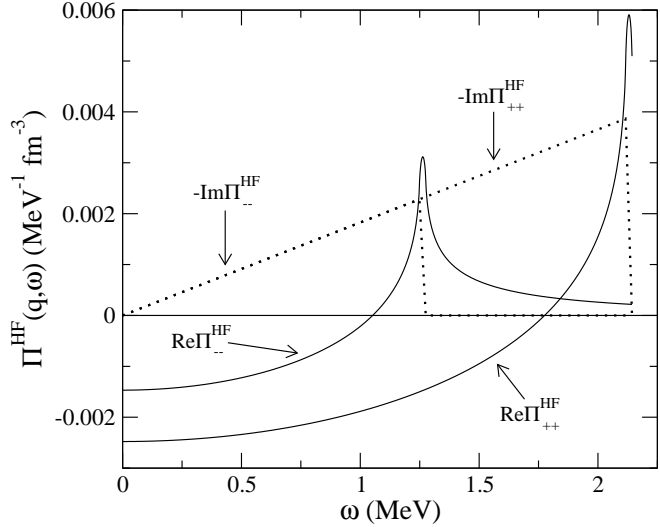}
\caption{\label{fig:Pizz_q005_Re0} Re$\Pi^{\textrm{HF}}_{\pm\pm}$ (solid) and
  -Im$\Pi^{\textrm{HF}}_{\pm\pm}$ (dot) as a function of $\omega$ at
  $q=5$~MeV/c. Note that Re$\Pi^{\textrm{HF}}_{\pm\pm}$ has a maximum at the
  upper limit of the corresponding response domain. }
\end{center}
\end{figure}

In order to understand the frequency behavior of the response at $q=5$ and
$q=50$~MeV/c it helps to keep in mind that 
\begin{itemize}
\item[a)] the HF (free) response in the broken, but not fully so, vacuum
  already displays two maxima, when the Pauli principle is active (namely for
  not too large $q$);
\item[b)] the RPA-HF framework conserves the energy-weighted sum rule even when
  the vacuum is broken (this non trivial result will be further discussed
  later).
\end{itemize}
Actually, the RPA-HF expression of the response reads
\begin{equation}
  -\frac{1}{\pi\rho}\textrm{Im}\Pi^{\textrm{RPA-HF,b}}_{zz} = 
    \frac{N_{\textrm{RPA-HF}}}{D_{\textrm{RPA-HF}}},
\end{equation}
being
\begin{equation}
  N_{\textrm{RPA-HF}} = -\frac{1}{\pi\rho} \left[ 
    \textrm{Im}(\Pi^{\textrm{HF}}_{++}+\Pi^{\textrm{HF}}_{--})
    (1-9V_1^2 u) + 6 V_1 v (1 + \frac{3}{2} V_1 
    \textrm{Re}(\Pi^{\textrm{HF}}_{++}+\Pi^{\textrm{HF}}_{--}) \right]
\end{equation}
and
\begin{equation}
  D_{\textrm{RPA-HF}} = 1 - 18 V_1^2 u + (9 V_1^2)^2 
    |\Pi^{\textrm{HF}}_{++}|^2 |\Pi^{\textrm{HF}}_{--}|^2,
\end{equation}
where
\begin{equation}
  u = \textrm{Re}\Pi^{\textrm{HF}}_{++} \textrm{Re}\Pi^{\textrm{HF}}_{--} - 
    \textrm{Im}\Pi^{\textrm{HF}}_{++} \textrm{Im}\Pi^{\textrm{HF}}_{--}
\end{equation}
and
\begin{equation}
  v = \textrm{Re}\Pi^{\textrm{HF}}_{++} \textrm{Im}\Pi^{\textrm{HF}}_{--} + 
    \textrm{Im}\Pi^{\textrm{HF}}_{++} \textrm{Re}\Pi^{\textrm{HF}}_{--}.
\end{equation}
In the high energy domain where only the spin up neutrons respond to the
external probe (namely, where $\textrm{Im}\Pi^{\textrm{HF}}_{--}=0$) the above
become 
\begin{equation}
  N_{\textrm{RPA-HF}} = -\frac{1}{\pi\rho} \textrm{Im}\Pi^{\textrm{HF}}_{++}
    \left[1 + 6 V_1 \textrm{Re}\Pi^{\textrm{HF}}_{--}(1 + 
    \frac{3}{2} V_1 \textrm{Re}\Pi^{\textrm{HF}}_{--}) \right]
\end{equation}
and
\begin{equation}
  D_{\textrm{RPA-HF}} = \left(1 - 9 V_1^2 
    \textrm{Re}\Pi^{\textrm{HF}}_{++} \textrm{Re}\Pi^{\textrm{HF}}_{--}
    \right)^2 + (9 V_1^2)^2 (\textrm{Im}\Pi^{\textrm{HF}}_{++})^2
    (\textrm{Re}\Pi^{\textrm{HF}}_{--})^2, 
\end{equation}
respectively, and since in this regime $\textrm{Re}\Pi^{\textrm{HF}}_{--}$ is a
rapidly decreasing function of the frequency (see
Fig.~\ref{fig:Pizz_q005_Re0}), it follows that the RPA-HF response at large
$\omega$ approaches the free one and thus the second peak displayed by the
latter still shows up. 

By contrast, in ring approximation one has the simpler expressions
\begin{equation}
  N_{\textrm{ring}} = -\frac{1}{\pi\rho}
    \textrm{Im}(\Pi^{\textrm{HF}}_{++}+\Pi^{\textrm{HF}}_{--}) 
\end{equation}
and
\begin{equation}
  D_{\textrm{ring}} = 1 - V_1 \textrm{Re}(\Pi^{\textrm{HF}}_{++} +
    \Pi^{\textrm{HF}}_{--}) + V_1^2 \left[ \textrm{Im}(\Pi^{\textrm{HF}}_{++} +
    \Pi^{\textrm{HF}}_{--}) \right]^2
\end{equation}
which, in the high-frequency regime, become
\begin{equation}
  N_{\textrm{ring}} = -\frac{1}{\pi\rho} \textrm{Im}\Pi^{\textrm{HF}}_{++}
\end{equation}
and
\begin{equation}
  D_{\textrm{ring}} = 1 - V_1 \textrm{Re}(\Pi^{\textrm{HF}}_{++} +
    \Pi^{\textrm{HF}}_{--}) + V_1^2 (\textrm{Im}\Pi^{\textrm{HF}}_{++})^2.
\end{equation}
From the last equation it appears that at large $\omega$ the ring response is
strongly damped.

\section{The system's transverse response}
\label{sec:xaxis}

In this Section we explore the system's response to a probe aligned in 
the direction orthogonal to the axis along which the spontaneous magnetization 
of the system occurs. For definitiveness we choose the probe to act in the 
$x$-direction. According to the general theory in a non-relativistic context
\cite{Nie76}, we expect here Goldstone modes to show up. 
Their number should not be less than the number of the broken generators 
of the continuous symmetry, provided that the Goldstone bosons of type II 
are counted twice. 
In the case we are presently investigating, the number of the broken generators
is provided by the dimensions of the coset $O(3)/O(2)$, where $O(3)$ is the
rotation group in three dimensions. This group leaves invariant the Hamiltonian
of our system of interacting neutrons, whereas $O(2)$ is the rotation group in
two dimensions and represents the surviving symmetry after the spontaneous
breaking has occurred. Hence in our case two generators are broken.
Accordingly this situation is compatible with the existence either of two 
Goldstone bosons of type I --- characterized by a dispersion relation linear in
the momentum --- or with the existence of one type II Goldstone boson ---
which has a dispersion relation quadratic in the momentum.

As we shall see, the latter is actually the occurring case for our system. 
This is hardly surprising since indeed type II Goldstone bosons are 
specific of a non-relativistic theory as exemplified by the rotational 
bands of atomic nuclei, whose energy depends quadratically on the 
angular momentum (nuclei are finite systems!) \cite{Boh75} and by the 
pairing additional and removal modes in superconducting nuclei \cite{Bar04}, 
whose energy depends quadratically on the number of pairs.
Clearly, in the two above mentioned examples the broken symmetries are the 
rotational and the global gauge ones, which are spontaneously broken in
deformed and superconducting nuclei, respectively.
It would be interesting to study a suitable relativistic generalization of our
system, since in that case the previously mentioned counting rule for the
number of Goldstone bosons no longer holds \cite{Low02}.

\begin{figure}
\begin{center}
\includegraphics[clip,height=5cm]{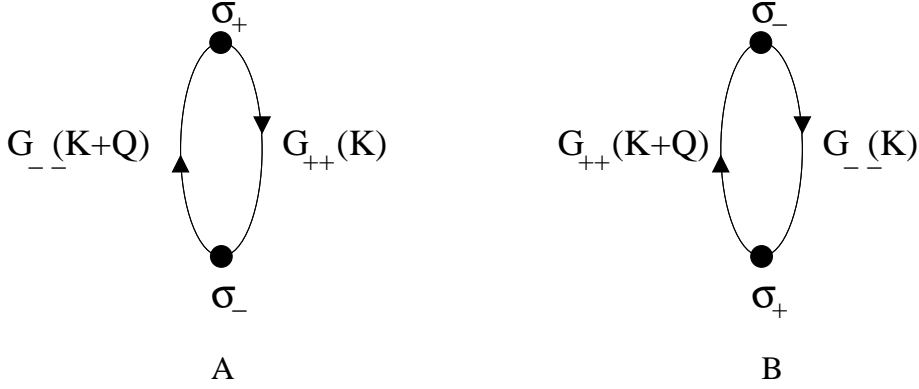}
\caption{\label{fig:pipiumeno}
  The diagrams corresponding to $\Pi^{\textrm{HF}}_{-+}(Q)$ (A) and 
  $\Pi^{\textrm{HF}}_{+-}(Q)$ (B).} 
\end{center}
\end{figure}

In searching for this Goldstone bosons we first ask: where do they live? To
answer this question we need to consider the transverse HF polarization
propagator 
\begin{eqnarray}
\label{piHFbxx}
  \Pi^{\textrm{HF},b}_{xx/yy}(Q) & = & - i
    \int\frac{{\mbox{d}^4}K}{(2\pi)^4} [G_{++}(K)G_{--}(K+Q)+
    G_{--}(K)G_{++}(K+Q)] \nonumber \\
  &=& \Pi^{\textrm{HF}}_{-+}(Q)+\Pi^{\textrm{HF}}_{+-}(Q),
\end{eqnarray}
where we have found it convenient to introduce the quantities
$\Pi^{\textrm{HF}}_{-+}$ and $\Pi^{\textrm{HF}}_{+-}$, whose vertices embody
the spin operators 
\begin{equation}
  \sigma_{\pm}=\frac{1}{2}(\sigma_x\,\pm\,i\sigma_y),
\end{equation}
as shown in Fig.~\ref{fig:pipiumeno}.
In the HF approximation the expressions for $\Pi_{-+}$ and $\Pi_{+-}$ are
easily deduced starting from the single particle propagators, already employed
in deducing the response to a longitudinal external probe, given in
Eqs.~(\ref{eq:GHF++}) and (\ref{eq:GHF--}).
One gets:
\begin{subequations}
\label{eq:Pipm}
\begin{eqnarray}
\label{eq:Pi+}
  \hskip -0.9cm \Pi^{\textrm{HF}}_{-+}(Q)&=&
    \int\frac{{\mbox{d}}\vec{k}}{(2\pi)^3}
    \left[\frac{\theta(|\vec{k}+\vec{q}|-k_F^-)\theta(k_F^+-k)}
    {\omega+\omega_{\vec{k}}^{+}-\omega_{\vec{k}+\vec{q}}^{-}
    +i\eta}-\frac{\theta(k_F^{-}-|\vec{k}+\vec{q}|)\theta(k-k_F^{+})}
    {\omega+\omega_{\vec{k}}^{+}-\omega_{\vec{k}+\vec{q}}^{-}-i\eta}\right] 
    \nonumber \\ \\\nonumber
\end{eqnarray}
and
\begin{eqnarray}
\label{eq:Pi-}
  \hskip -0.9cm \Pi^{\textrm{HF}}_{+-}(Q)&=&
    \int\frac{{\mbox{d}}\vec{k}}{(2\pi)^3}
    \left[\frac{\theta(|\vec{k}+\vec{q}|-k_F^+)\theta(k_F^- -k)}
    {\omega+\omega_{\vec{k}}^{-}-\omega_{\vec{k}+\vec{q}}^{+}
    +i\eta}-\frac{\theta(k_F^{+}-|\vec{k}+\vec{q}|)\theta(k-k_F^{-})}
    {\omega+\omega_{\vec{k}}^{-}-\omega_{\vec{k}+\vec{q}}^{+}-i\eta}\right].
    \nonumber \\
\end{eqnarray}
\end{subequations}
From the above formulas, the response region of the infinite, homogeneous 
neutron's system in the $(\omega,q)$ plane to a spin-flipping probe
($\sigma_\pm$) is deduced by searching for the region where, e.~g., 
$\Pi^{\textrm{HF}}_{-+}(Q)$ develops an imaginary part. For this purpose we
write
\begin{equation}
  \textrm{Im}\Pi^{\textrm{HF}}_{-+}(Q)=\textrm{Im}\Pi^{a}_{-+}(Q)+
    \textrm{Im}\Pi^{b}_{-+}(Q),
\label{eq:impi+}
\end{equation}
being
\begin{subequations}
\begin{eqnarray}\label{pipiua}
  \textrm{Im}\Pi^{a}_{-+}(Q)&=&\int\frac{{\mbox{d}}\vec{k}}{(2\pi)^3}
    \theta(|\vec{k}+\vec{q}|-k_F^-)\theta(k_F^+-k)(-\pi)
    \delta(\omega+\omega_{\vec{k}}^{+}-\omega_{\vec{k}+\vec{q}}^{-})
    \nonumber \\
 \end{eqnarray}
and
\begin{eqnarray}\label{pipiub}   
  \textrm{Im}\Pi^{b}_{-+}(Q)&=&\int\frac{{\mbox{d}}\vec{k}}{(2\pi)^3}
    \theta(k_F^{-}-|\vec{k}+\vec{q}|)\theta(k-k_F^{+})(-\pi)
    \delta(\omega+\omega_{\vec{k}}^{+}-\omega_{\vec{k}+\vec{q}}^{-}).
    \nonumber \\
\end{eqnarray}
\end{subequations}
Hence, the first contribution to the imaginary part of $\Pi_{-+}^{\textrm{HF}}$
(namely Im$\Pi^{a}_{-+}(Q)$) lives in the domain
\begin{subequations}
\label{eq:Pi-+domain}
\begin{equation}
\label{eq:Pi-+adomain}
  \frac{q^2}{2m}-\frac{k_F^+q}{m}+\Delta\omega
    <\omega<\frac{q^2}{2m}+\frac{k_F^+q}{m}+\Delta\omega, 
\end{equation}
while the second one (namely Im$\Pi^{b}_{+}(Q)$) lives in the domain
\begin{equation}
\label{eq:Pi-+bdomain}
  -\frac{q^2}{2m}-\frac{k_F^-q}{m}+\Delta\omega
    <\omega<-\frac{q^2}{2m}+\frac{k_F^-q}{m}+\Delta\omega.
\end{equation}
\end{subequations}
In the above we have set
\begin{equation}
\label{eq:Deltaomega}
  \Delta\omega=-\frac{V_1}{2\pi^2}\left[(k_F^{+})^3-(k_F^{-})^3\right]
\end{equation}
and in our derivation we have exploited Eq.~(\ref{eq:zerorange}) and the
relation 
\begin{equation}
  \omega_{\vec{k}}^{+}-\omega_{\vec{k}+\vec{q}}^{-}=
    \omega_{\vec{k}}-\omega_{\vec{k}+\vec{q}}-\Delta\omega=
    \omega_{\vec{k}}-\omega_{\vec{k}+\vec{q}}-
    \frac{1}{2m}((k_F^{+})^2-(k_F^{-})^2),
\end{equation}
the last equality being valid for 
$V_{1,c}^{\textrm{lower}}\le V_1 \le V_{1,c}^{\textrm{upper}}$.
For definitiveness, in the following we shall always choose $k_F^+\ge
k_F^-$,which implies $\Delta\omega\ge0$.

\begin{figure}
\begin{center}
\includegraphics[clip,height=7cm]{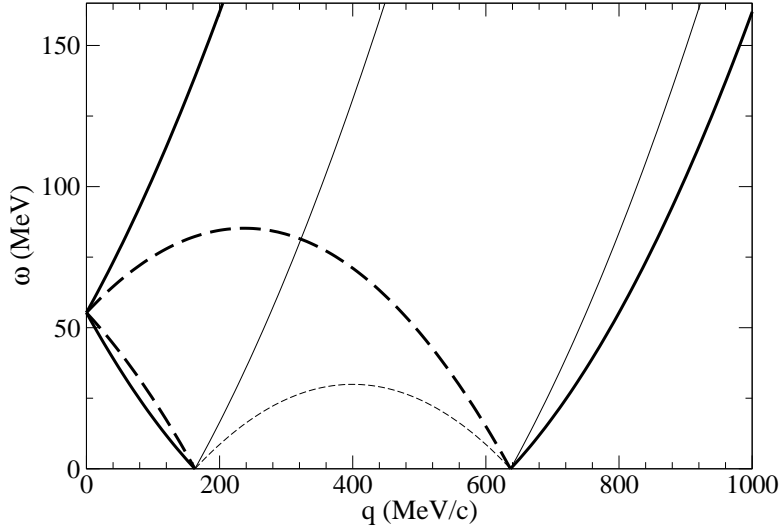}
\caption{\label{fig:rispxx}
  The response region associated to $\Pi_{-+}^{\textrm{HF}}$ (heavy lines) and
  to $\Pi_{+-}^{\textrm{HF}}$ (light lines). The solid and dashed lines
  correspond to the first and second term in which we have split the imaginary
  part of $\Pi^{\textrm{HF}}$, respectively (see Eq.~(\protect\ref{eq:impi+})).
  The values for $k_F^+$ and $k_F^-$ are the same as in
  Fig.~\protect\ref{fig:resp_region}.}
\end{center}
\end{figure}
The response region of $\Pi^{\textrm{HF}}_{+-}(Q)$ can be derived along the
same lines, yielding, instead of Eq.~(\ref{eq:Pi-+domain}), the following
expressions:
\begin{subequations}
\begin{eqnarray}
  \frac{q^2}{2m}-\frac{k_F^-q}{m}-\Delta\omega<&\omega&<
    \frac{q^2}{2m}+\frac{k_F^-q}{m}-\Delta\omega \\
  -\frac{q^2}{2m}-\frac{k_F^-q}{m}-\Delta\omega<&\omega&<
    -\frac{q^2}{2m}+\frac{k_F^-q}{m}-\Delta\omega.
\end{eqnarray}
\end{subequations}
It is of importance to observe that the response region related to 
$\Pi^{\textrm{HF}}_{-+}$ ($\Pi^{\textrm{HF}}_{+-}$) is shifted with respect to
the symmetric case upward (downward) by an amount $\Delta\omega$ that directly
reflects the size of the spontaneous breaking of the vacuum.
The response regions for $\Pi_{-+}^{\textrm{HF}}$ and $\Pi_{+-}^{\textrm{HF}}$
are displayed in Fig.~\ref{fig:rispxx}. 

Concerning the response function, it is remarkable that the following symmetry
relation holds valid: $\Pi_{-+}^{\textrm{HF}}(\vec{q},\omega)=
\Pi_{+-}^{\textrm{HF}}(\vec{q},-\omega)$, as one can see by comparing
Eqs.~(\ref{eq:Pi+}) and (\ref{eq:Pi-}).
Furthermore, note that, at variance with the symmetric vacuum case, now, for 
$\omega>0$, also the second piece on the right hand side of Eqs.~(\ref{eq:Pi+})
and (\ref{eq:Pi-}) contributes to the system's response, the more so, the
smaller $q$ is.  
To complete our analysis we quote in Appendix~\ref{app:im_re_pi_+-} the real
and the imaginary part of $\Pi_{-+}^{\textrm{HF}}$ and
$\Pi_{+-}^{\textrm{HF}}$. 

It should finally  be stated that similar results have been obtained in the
context of asymmetric nuclear matter \cite{Alb89,Tak92}.

\begin{figure}
\begin{center}
\includegraphics[clip,height=6cm]{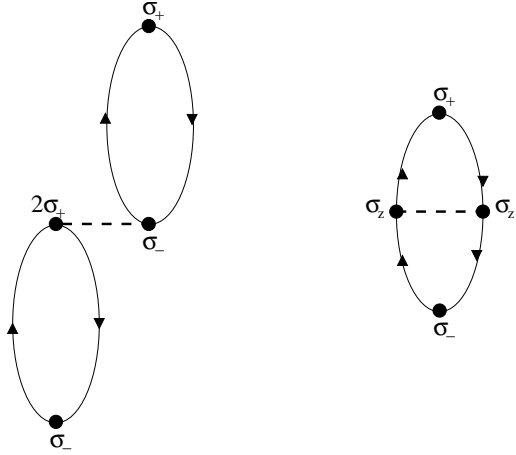}
\caption{\label{fig:pipiurpa}
  The first order direct and exchange terms for $\Pi^{\textrm{RPA}}_{-+}(Q)$.} 
\end{center}
\end{figure}

We turn now to discuss the RPA equations for $\Pi_{-+}$ and $\Pi_{+-}$. 
The basic ingredients required for their deduction are the \emph{first order}
direct and exchange diagrams displayed (for $\Pi_{+}$) in 
Fig.~\ref{fig:pipiurpa}. In our case of a zero-range interaction one finds
\begin{equation}
  \Pi_{\mp\pm}^{\textrm{(1)dir}}(Q) = 2V_1\Pi_{\mp\pm}^{\textrm{HF}}(Q)
    \Pi_{\mp\pm}^{\textrm{HF}}(Q)
\end{equation}
for the direct term and
\begin{equation}
  \Pi_{\mp\pm}^{\textrm{(1)ex}}(Q) = V_1\Pi_{\mp\pm}^{\textrm{HF}}(Q)
    \Pi_{\mp\pm}^{\textrm{HF}}(Q)
\end{equation}
for the exchange diagram.
Hence the RPA series (which accounts for both contributions) can be
easily resummed, leading to
\begin{equation}
\label{resummed}
  \Pi_{\mp\pm}^{\textrm{RPA-HF}}(Q) = 
    \frac{\Pi_{\mp\pm}^{\textrm{HF}}(Q)}{1-3 V_1\Pi_{\mp\pm}^{\textrm{HF}}(Q)}.
\end{equation}  
To find the dispersion relation of the Goldstone bosons we search for the poles
(if any) of the expression 
\begin{equation}
\label{disppix}
  \Pi_{xx}^{\textrm{RPA-HF}}(\vec{q},\omega) = 
    \frac{\Pi_{-+}^{\textrm{HF}}(\vec{q},\omega)
    +\Pi_{+-}^{\textrm{HF}}(\vec{q},\omega)
    -6V_1\Pi_{-+}^{\textrm{HF}}(\vec{q},\omega)
    \Pi_{+-}^{\textrm{HF}}(\vec{q},\omega)}
    {[1-3 V_1\Pi_{-+}^{\textrm{HF}}(\vec{q},\omega)]
    [1-3 V_1\Pi_{+-}^{\textrm{HF}}(\vec{q},\omega)]}
\end{equation}
for positive real $\omega$.\\
From the numerical analysis we have found that of the two factors appearing 
in the denominator of Eq.~(\ref{disppix}) only the first one (since
$k_F^+>k_F^-$) vanishes for just one real and positive value of $\omega$ at a
given $q$. 
\begin{figure}
\begin{center}
\includegraphics[clip,width=\textwidth]{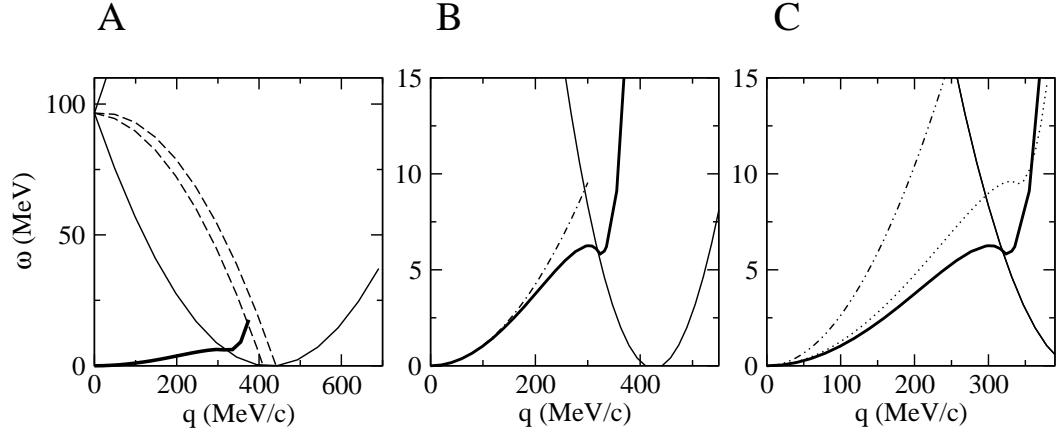}
\caption{\label{fig:dispersione} The dispersion relation of the Goldstone boson
  for $k_F^+=426.01$~MeV/c (heavy solid lines), which corresponds to
  $V_1=-189.25$~MeV~fm$^3$. 
  Also displayed are the response regions: the light solid
  and dashed lines correspond to Eqs.~(\protect\ref{eq:Pi-+adomain}) and
  (\protect\ref{eq:Pi-+bdomain}), respectively. In panel A one can appreciate
  how tiny the energy of the Goldstone boson is; in panel B, which enlarges
  panel A, one can assess the domain of validity of the parabolic dispersion
  relation of the Goldstone mode (dot-dashed line); in panel C the Goldstone
  mode is displayed for three different values of the interaction strength,
  namely $V_1=-189.25$ (solid), $-200$ (dot) and $-300$~MeV~fm$^3$
  (dot-dot-dash).} 
\end{center} 
\end{figure} 
In Fig.~\ref{fig:dispersione} we display the solution of the equation
\begin{equation}
\label{dispe}
  [1-3 V_1\textrm{Re} \Pi_{-+}^{\textrm{HF}}(\vec{q},\omega)]=0,
\end{equation}
which we expect to yield the dispersion relation of the Goldstone boson, 
should the RPA be a trustworthy theory for our many-body system. This turns 
out indeed to be the case, since for small $q$ and $\omega$ the solution of 
Eq.~(\ref{dispe}) can be analytically expressed through the expansion of 
Re$\Pi_{-+}^{\textrm{HF}}(\vec{q},\omega)$, which reads
\begin{eqnarray} 
  \lim_{q\to0}\textrm{Re} \Pi_{-+}^{\textrm{HF}}(\vec{q},\omega)& = &
    \frac{mk_F^+}{4\pi^2}\frac{1}{\nu-\Delta\nu}
    \left[\frac{2}{3}(1-\xi^3)+\frac{1}{3}\frac{Q^2}{\nu-\Delta\nu}(1+\xi^3)
    \right.\nonumber\\
  & + &\left.\frac{2}{15}\frac{Q^2}{(\nu-\Delta\nu)^2}(1-\xi^5)\right],
\end{eqnarray}
where
\begin{equation} 
  Q=\frac{q}{k_F^+},\ \xi=\frac{k_F^-}{k_F^+}, \ 
    \nu=\frac{m\omega}{(k_F^+)^2} \ \ \textrm{and} \ \ \Delta\nu=
    \frac{-mV_1}{2\pi^2(k_F^+)^2}[(k_F^+)^3-(k_F^-)^3].
    \nonumber
\end{equation}
From the above one gets the following dispersion relation, valid for small
values of $q$:
\begin{equation}
\label{parstar}
  \omega=\frac{q^2}{2m^{\star}}
\end{equation}
with 
\begin{equation}
  m^{\star}=\frac{5(1+\xi)(1+\xi+\xi^2)}{(1-\xi)(1+3\xi+\xi^2)}m.
\end{equation}
From Fig.~\ref{fig:dispersione} (panels B and C) it appears that the expression
(\ref{parstar}) actually remains valid over a substantial range of momenta. 
Thus, the solution of Eq.~(\ref{dispe}) truly corresponds to a type II 
Goldstone boson as it should.
Microscopically this mode behaves like a particle-hole excitation (the
particle being quite heavy).  
Physically it can be viewed as a twisting of the local spin orientation as the
collective wave passes through the system \cite{For75}.

Furthermore, and remarkably, it turns out that for
$V_1>V_{1,c}^{\textrm{upper}}$ the Goldstone mode continues to exist with a
dispersion relation that is parabolic over a range of momenta becoming larger
as $V_1$ increases. For $V_{1,c}^{\textrm{lower}}\le V_1\le
V_{1,c}^{\textrm{upper}}$ the Goldstone mode displays instead an anomalous
behavior: in fact, in this range of couplings, in correspondence to a specific
momentum, the collective mode is characterized by a vanishing group velocity.

It is worth comparing the dispersion relation (\ref{parstar}) with the 
formula for the energy levels of a rotational band. They are identical  
providing one replaces the momentum with the quantized angular momentum (of
course, both quantities should not be too large) and the mass with the moment
of inertia. From Fig.~\ref{fig:dispersione} (panel A) it is clearly apparent
how tiny the energy of the Goldstone boson is, just as  it happens for the
rotational bands in nuclear and molecular physics.
\begin{figure}
\begin{center}
\includegraphics[clip,width=\textwidth]{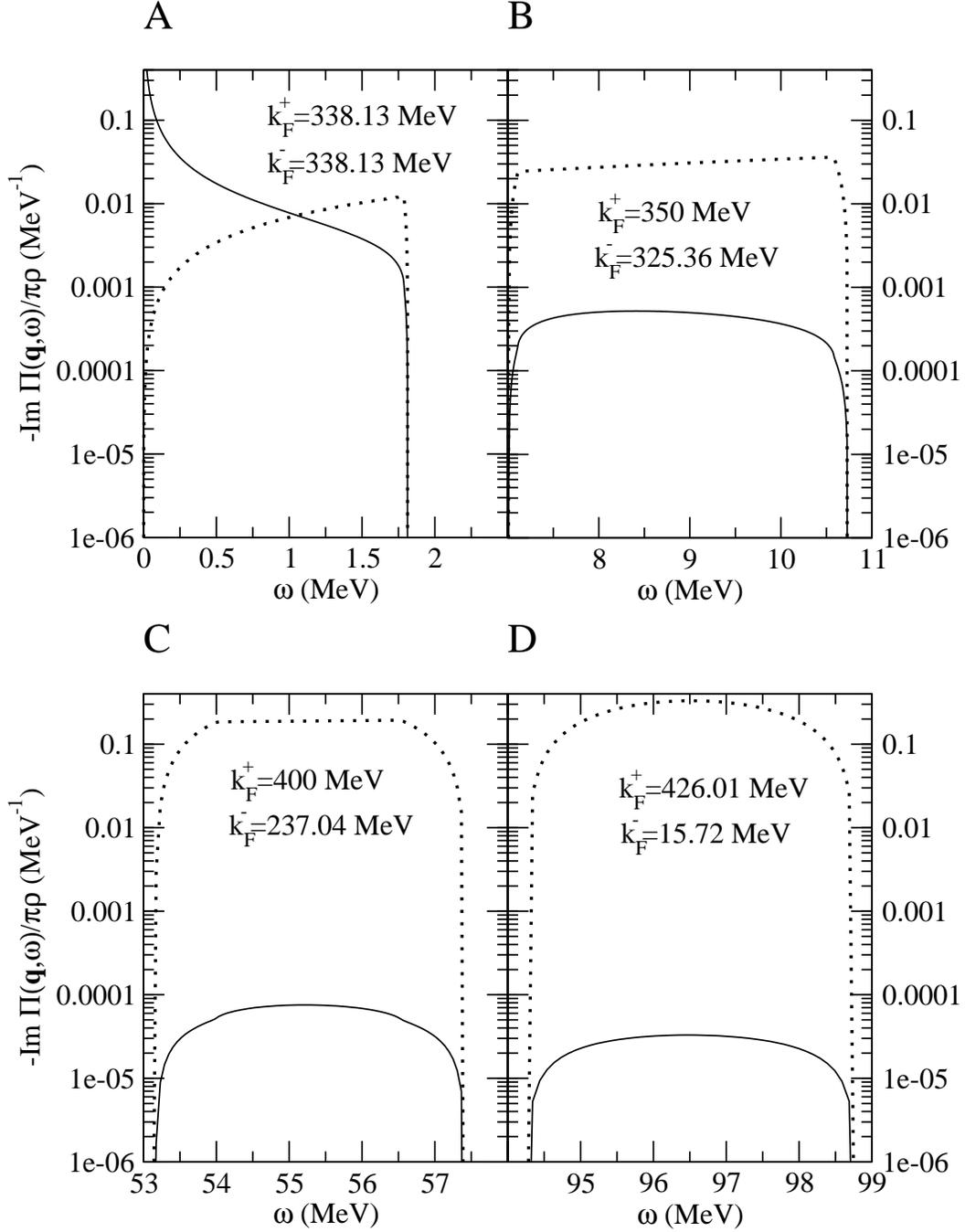}
\caption{\label{fig:Pixx_q005} The response of an infinite neutron's 
  system to a $x$-aligned probe for $q=5$ MeV/c. Panel (A) refers to the case 
  of a symmetric vacuum ($k_F=k_{F}^{+}=k_{F}^{-}$), panels (B) and (C) to 
  a partially aligned broken vacuum, panel (D) to a totally broken, fully 
  aligned vacuum. Dotted line: HF response, solid line: RPA-HF.} 
\end{center}
\vskip 1cm
\end{figure} 
\begin{figure}
\begin{center}
\includegraphics[clip,width=\textwidth]{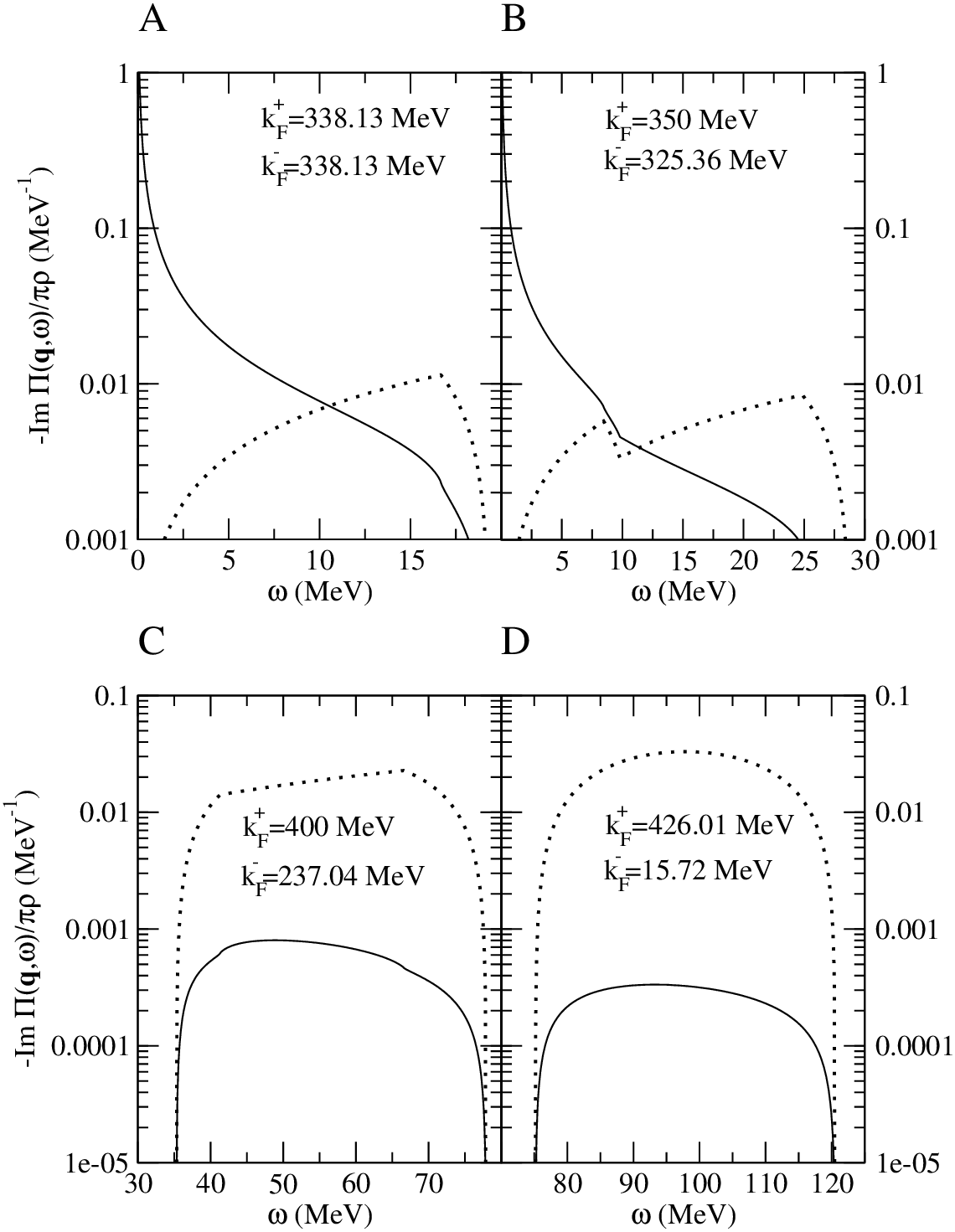}
\caption{\label{fig:Pixx_q050} The same as in Fig.~\protect\ref{fig:Pixx_q005}
   but for $q=50$ MeV/c. }
\end{center}
\vskip 3cm
\end{figure} 
\begin{figure}
\begin{center}
\includegraphics[clip,width=\textwidth]{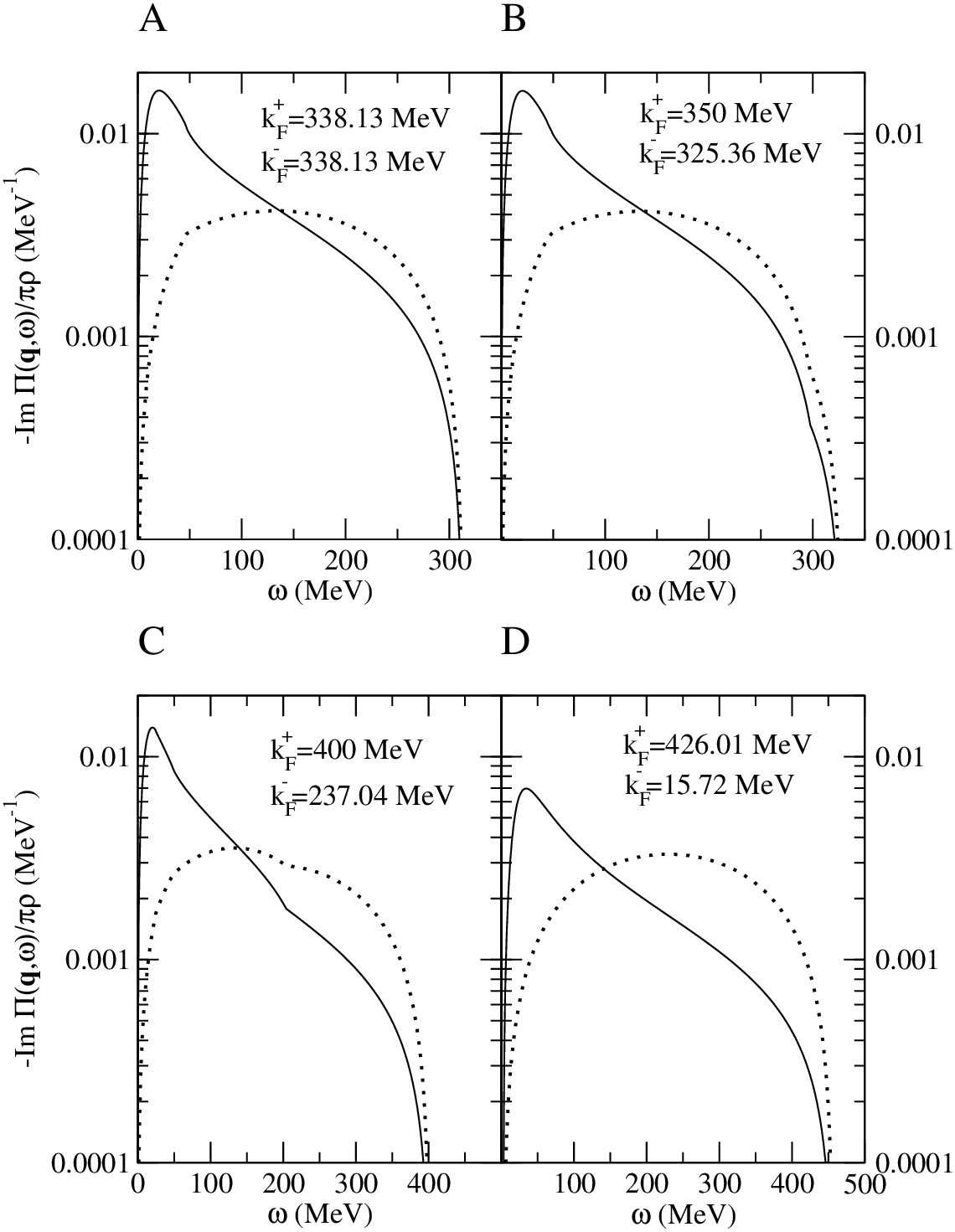}
\caption{\label{fig:Pixx_q500} As in Fig.~\protect\ref{fig:Pixx_q005}
  but for $q=500$ MeV/c. } 
\end{center}
\vskip 3cm
\end{figure} 

In concluding this Section we display in Figs.~\ref{fig:Pixx_q005}, 
\ref{fig:Pixx_q050} and \ref{fig:Pixx_q500} the continuum response 
of the system to an $x$-aligned probe for the same momenta and vacua of 
Figs.~\ref{fig:Pizz_q005}, \ref{fig:Pizz_q050} and \ref{fig:Pizz_q500}, where
the response to a $z$-aligned probe was considered. We notice that
\begin{itemize}
\item[i)] for a symmetric vacuum the energy-weighted sum rule is patently 
obeyed;
\item[ii)] the more the vacuum is broken, the more depleted the particle-hole 
continuum is;
\item[iii)] in accord with ii) the more the vacuum is broken, the stronger the 
Goldstone boson becomes. This item will be quantitatively addressed in the next
section in the sum rule framework.
\end{itemize}

\section{The moments of the response function}
\label{sec:moments}

In this Section we investigate the non-energy-weighted ($S_0$) and the
energy-weighted ($S_1$) sum rules, exploring their behavior when spontaneous
symmetry breaking occurs in the vacuum.

Concerning $S_1$, it is well-known \cite{Rin80} that it is given by the
following expression 
\begin{equation}
\label{eq:S1}
  S_1(q) = \int_0^\infty\mbox{d}\omega \, \omega
    \frac{-\textrm{Im}\Pi(\vec{q},\omega)}{\pi\rho} 
    = \frac{1}{2} \langle0|[\widehat{O},[\hat{H},\widehat{O}]]|0\rangle.
\end{equation}
For the density response of a non-interacting gas of fermions of mass $m$ the
above is indeed fulfilled and yields
\begin{equation}
  S_1^{\textrm{free}}(q) = \int_0^\infty\mbox{d}\omega \, \omega
    \frac{-\textrm{Im}\Pi^0(\vec{q},\omega)}{\pi\rho} 
    = \frac{q^2}{2m}.
\end{equation}
In general, however, Eq.~(\ref{eq:S1}) is violated by most of the many-body
frameworks, the remarkable exception being the RPA-HF theory.
In fact, the Thouless theorem \cite{Tho61} states that if the system's response
is computed in RPA-HF and the expectation value on the right hand side of
Eq.~(\ref{eq:S1}) is taken in the HF ground state, then the sum rule is
fulfilled. We have indeed verified it by computing numerically with very good
accuracy the left hand side of Eq.~(\ref{eq:S1}) and by working out the HF
expectation value of the double commutator in the same equation --- which of
course yields $q^2/2m$, if one employs the interaction (\ref{Vdelta})
and the vertex (\ref{ext_probe}).

Remarkably, even when the vacuum is broken $S_1$ keeps the above value, as it
can be inferred from the results reported in Tables~\ref{tabella_q_5},
\ref{tabella_q_50} and \ref{tabella_q_500}. Note that this outcome could also
be proved along the lines followed in Ref.~\cite{Tak90} in the case of a
symmetric vacuum. Conceptually, it relates to the very meaning of $S_1$, namely
of expressing the particle number conservation (or the global gauge invariance)
of the theoretical framework and in the non-symmetric vacuum it is the
rotational --- and not the gauge --- invariance to be broken. 

\begin{table}[!htb]
\begin{tabular}{llllll}
  $k_{F}^{+}$ & $S_0^{\textrm{HF}}$ & $S_0^{\textrm{RPA-HF}}$ & 
    $S_1^{\textrm{HF}}$ & $S_1^{\textrm{RPA-HF}}$ & $\Delta\omega$ \\
  \hline
  338.130028 & 0.0111 & (0.0819+0) & 0.013312 & (0.013312+0) & 0\\
  350 & 0.1091 & (0.00152+0.10754) & 0.9795 & (0.013295+0.000017) & 8.860\\
  400 & 0.65549 & (0.000228+0.65527) & 36.248 & (0.012594+0.000718) & 55.279\\
  426.01 & 1.000 & (0.000111+0.99979) & 96.510 & (0.010664+0.002686) & 96.505\\
\end{tabular}
\caption{\label{tabella_q_5} The non-energy-weighted and energy-weighted sum
    rules at $q=5$ MeV/c, corresponding to $q^2/2m=0.0133120$ MeV.
    The Fermi momentum $k_F^+=338.130028$~MeV/c corresponds to a symmetric
    vacuum, whereas $k_F^+=426.01$~MeV/c corresponds to an almost completely
    broken vacuum (ground state fully aligned in spin space).
    In the columns associated with the RPA-HF theory, the first figure
    represents the contribution to $S_0$ and $S_1$, respectively, arising from
    the particle-hole continuum; the second figure the one arising from the
    collective Goldstone mode. }
\vskip 1cm
\begin{tabular}{llllll}
  $k_{F}^{+}$ & $S_0^{\textrm{HF}}$ & $S_0^{\textrm{RPA-HF}}$ & 
    $S_1^{\textrm{HF}}$ & $S_1^{\textrm{RPA-HF}}$ & $\Delta\omega$ \\
  \hline
  338.130028 & 0.1107 & (0.5842+0) & 1.33120 & (1.33120+0) & 0\\
  350 & 0.1374 & (0.5340+0) & 2.29744 & (1.33120+0) & 8.860\\
  400 & 0.6555 & (0.0233+0.6322) & 37.566 & (1.26414+0.06706) & 55.279\\
  426.01 & 0.9999 & (0.0111+0.9888) & 97.827 & (1.07098+0.26025) & 96.505\\
\end{tabular}
\caption{\label{tabella_q_50} The same as in Table~\protect\ref{tabella_q_5} 
  but for $q=50$ MeV/c, corresponding to $q^2/2m=1.33120$ MeV.}
\vskip 1cm
\begin{tabular}{llllll}
  $k_{F}^{+}$ & $S_0^{\textrm{HF}}$ & $S_0^{\textrm{RPA-HF}}$ & 
    $S_1^{\textrm{HF}}$ & $S_1^{\textrm{RPA-HF}}$ & $\Delta\omega$ \\
  \hline
  338.130028 & 0.907 & 1.528 & 133.120 & 133.120 &0\\
  350 & 0.908 & 1.524 & 134.087 & 133.120 & 8.860\\
  400 & 0.951 & 1.359 & 169.356 & 133.120 & 55.279\\
  426.01 & 1.000 & 1.000 & 229.616 & 133.120 & 96.505\\
\end{tabular}
\caption{\label{tabella_q_500} The same as in Table~\protect\ref{tabella_q_5} 
  but for $q=500$ MeV/c, corresponding to $q^2/2m=133.120$ MeV.}
\vskip 2.5cm
\end{table}

In this instance, at variance with the situation where the probe acts in the
direction of the spontaneous magnetization, when the spin-flipping probe is
directed orthogonally to the latter, $S_1$ is contributed to not only by the
particle-hole continuum, but by the collective Goldstone mode as well.
Indeed, from Eq.~(\ref{resummed}) one has
\begin{equation}
  \textrm{Im}\Pi_{-+}^{\textrm{RPA-HF}}(Q) =
    \frac{\textrm{Im}\Pi_{-+}^{\textrm{HF}}(Q)}
    {\left[ 1-3V_1\textrm{Re}\Pi_{-+}^{\textrm{HF}}(Q) \right]^2 +
    9 V_1^2 \left[\textrm{Im}\Pi_{-+}^{\textrm{HF}}(Q) \right]^2 },
\end{equation}
which, in the region where $\textrm{Im}\Pi_{-+}^{\textrm{HF}}(Q)$ vanishes,
yields 
\begin{eqnarray}
  \textrm{Im}\Pi_{-+}^{\textrm{RPA-HF}}(Q) &=& \pi
    \textrm{Re}\Pi_{-+}^{\textrm{HF}}(Q) 
    \delta[1-3V_1\textrm{Re}\Pi_{-+}^{\textrm{HF}}(Q)] \nonumber \\
  &=& \frac{\pi}{9V_1^2} \frac{1}{\left|
    \frac{\displaystyle\partial\textrm{Re}\Pi_{-+}^{\textrm{HF}}(Q)}
    {\displaystyle\partial\omega}
    \right|_{\omega=\omega_+(q)}} \delta[\omega-\omega_+(q)], 
\end{eqnarray}
with $\omega_+(q)$ the solution of Eq.~(\ref{dispe}), that is the Goldstone
boson dispersion relation.
\begin{figure}
\begin{center}
\includegraphics[clip,width=0.7\textwidth]{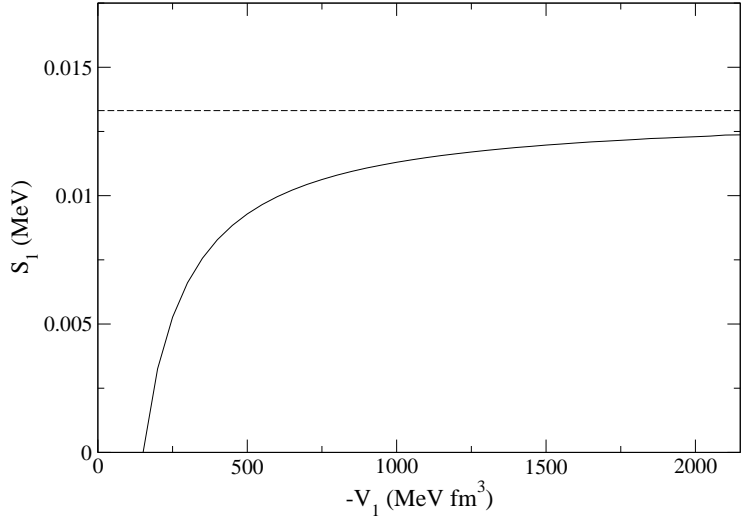}
\caption{\label{fig:oltre-v-crit} Contribution of the Goldstone mode to the
  energy-weighted sum rule as a function of the interaction strength for
  $q=5$~MeV/c (solid line); the curve starts at
  $V_{1,c}^{\textrm{lower}}=-159.16$~MeV~fm$^3$. The dashed line represents the
  saturation value $q^2/2m=0.0133120$~MeV. }
\end{center}
\end{figure} 

Actually, this contribution grows with the amount of symmetry breaking in the
vacuum, which, in turn, grows with the strength of the force $V_1$. 
For $V_1=V_{1,c}^{\textrm{upper}}$, namely when the vacuum is fully aligned in
spin-space, the Goldstone mode accounts for roughly 25\% of the energy-weighted
sum rule, as it can be deduced from the figures reported in
Tables~\ref{tabella_q_5} and \ref{tabella_q_50}. For still larger values of
$V_1$, this amounts keeps increasing until, for $V_1\to\infty$, it exhausts
the sum rule, as it is seen in Fig.~\ref{fig:oltre-v-crit}, where the Goldstone
mode contribution to $S_1$ is displayed versus $V_1$. Of course, for large
values of $q$ the Goldstone boson no longer exists (Table~\ref{tabella_q_500}).

Concerning the non-energy-weighted sum rule in a symmetric vacuum and for a
non-interacting system of fermions one has the well-known result
\begin{eqnarray}
  S_0(q) &=& \frac{3}{4} \frac{q}{k_F} \left[1-\frac{1}{12}\left(
    \frac{q}{k_F}\right)^2\right] \phantom{1} \qquad q\le2k_F \nonumber \\
         &=& 1 \phantom{\frac{3}{4} \frac{q}{k_F} \left[1-\frac{1}{12}\left(
    \frac{q}{k_F}\right)^2\right]} \qquad q>2k_F,
\end{eqnarray}
which is conserved in the HF theory, but not in RPA or RPA-HF.
When the vacuum is spontaneously broken by our ferromagnetic force, one finds
that the impact of the Pauli correlations on $S_0$ is lowered with respect to
the symmetric vacuum case and decreases as the amount of the symmetry breaking
grows, as it is apparent from Tables~\ref{tabella_q_5}, \ref{tabella_q_50} and
\ref{tabella_q_500}. In particular, for a fully broken vacuum, $S_0$ is just
$1$ for any $q$ --- that is the value occurring in the symmetric vacuum for
$q\ge2k_F$ --- both when our system is explored in the longitudinal or in the
transverse direction by a spin-dependent probe: the system's constituents no
longer feel the Pauli principle, as it should be expected since in the fully
magnetized case only one species of particles is present.

The reduced influence of the Pauli principle, when the system in only partially
aligned, with respect to the situation occurring in the symmetric vacuum, can
be exploited to investigate (using a spin-flipping probe) how the collectivity
of the Goldstone mode is affected by the degree of spontaneous symmetry
breaking of the vacuum. In fact, in Tables~\ref{tabella_q_5} and 
\ref{tabella_q_50} one sees that the less effective the Pauli correlations
are, the more collective the Goldstone boson is.

\section{Conclusions}
\label{sec:concl}

In the present study we have dealt with an infinite, non-relativistic,
homogeneous system of neutrons interacting through a simple ferromagnetic force
of Heisenberg type, our aim being to discuss general aspects of the spontaneous
symmetry breaking associated with a quantum phase transition.
Although this theme has been much addressed in the past for generic spin $1/2$
fermions (see, e.~g., Ref.~\cite{Hua98}), still we felt it useful to analyze in
more detail the dependence upon the interaction range of the critical values of
the coupling and the excitation spectrum of a system undergoing a quantum phase
transition. 

Concerning the latter, we like to remind that the symmetry breaking taking
place in our system, namely the onset of a permanent, spontaneous
magnetization, stems from a well identified physical source, a situation very
different, for example, from the electroweak symmetry breaking in the standard
model, where the responsible for such an occurrence, namely the Higgs boson, is
still searched for three decades after having been conjectured. 
Indeed,  in our system --- just like in a ferromagnet the electromagnetic
interaction between neighboring atoms lines up their spins parallel to each
other --- the force between neighboring neutrons produces the same effect,
provided that the strength of the interaction exceeds some critical value,
$V_{1,c}^{\textrm{lower}}$. 
How close the neutrons should be for the phase transition to occur or,
equivalently, how the range of the interaction affects the value of
$V_{1,c}^{\textrm{lower}}$? We have answered this question both numerically and
analytically, finding out, as expected, that the longer the range, the weaker
the strength of the coupling should be in order to induce the spontaneous
breaking of a symmetry. 
A further natural question to ask is how large $V_1$ should be for the system
to become fully magnetized. Also for this problem we provide a numerical and
an analytic answer, again confirming the above referred to correlation between
strength and range.

We have next analyzed the spectrum of the system (or, equivalently, the
response functions to external probes acting on the spin of the constituents),
following its evolution with the amount of spontaneous symmetry breaking
occurring in the ground state.
Probing the system along the direction of the spontaneous magnetization, the
chief feature of the response function relates to the occurrence of two
distinct peaks, which reflect both the action of the ferromagnetic force and of
the Pauli principle. Indeed, the two peaks merge at large transferred momenta,
where the latter disappears.

In the direction orthogonal to the magnetization, on the other hand, a
collective Goldstone mode shows up as required by the theory.
Its dispersion relation is indeed parabolic, as it should, over a momentum
range that increases as the coupling strength $V_1$ increases.
Actually, for $V_1\to\infty$, the Goldstone boson exhausts the energy-weighted
sum rule and its dispersion relation becomes a perfect parabola.
Notably, for a strength $V_1$ intermediate between the two critical values
$V_{1,c}^{\textrm{lower}}$ and $V_{1,c}^{\textrm{upper}}$, the Goldstone boson
dispersion relation displays an anomalous non-parabolic behavior, entailing the
existence of a wavelength associated with a vanishing group velocity of the
collective mode. We conjecture this to be a distinctive signature of the
Goldstone boson in the non-relativistic regime.

In spite of the simplicity of our interaction, which has of course no pretense
of being realistic, it appears that our research bears significance for the
physics of the neutron stars, since it explores the extension of the many-body
response theory to the situation associated with a broken vacuum in a spin
space, which is required for the assessment of the magnetic field that neutron
stars host. A lot of work has actually been lately done on this issue: 
interestingly, it appears that simple effective interactions ---
such as the Skyrme ones --- give indeed rise to a phase transition of second
kind \cite{Sto03,Isa04}, whereas more microscopic many-body approaches --- such
as the Brueckner-Hartree-Fock formalism \cite{Vid02} or quantum simulations
\cite{Fan01,Sar03} --- give no indication of a quantum phase transition. 
Generally speaking, this striking difference can be related to the different
predictions these models give for the particle-hole spin interaction at neutron
star densities: attractive in the Skyrme models and repulsive in calculations
based on realistic nucleon-nucleon potentials \cite{Red99}.
Unfortunately, at present there are no direct phenomenological constraints on
this component of the effective nuclear interaction at densities relevant for
the neutron stars.

In connection with the possible application of the concept of spontaneous
symmetry breaking to the physics of the atomic nuclei, it should be of course
realized that in finite systems the notion of phase transition applies only
approximately. Yet, one is naturally lead to think of heavy nuclei, where the
isospin symmetry is broken, leading to ground state configurations quite
similar to the ones we have been considering in the spin space.  
However, in isospace the symmetry is broken explicitly (by the Coulomb force)
and not spontaneously (as in our case). 
Thus, although for heavy nuclei the response region of the system in the
$(\omega,q)$ plane turns out to be very similar in both cases \cite{Alb89}, in
nuclear matter no collective Goldstone boson should show up in the small
$\omega$, small $q$ corner of the $(\omega,q)$ plane. 
Furthermore, in the isospin particle-hole channel --- the one appropriate to
the present considerations --- no attraction seems to exist.

\section*{Appendices}
\appendix

\section{Probing the system in the symmetric direction: the RPA matrix
  elements} 
\label{app:rpazz}

\begin{figure}
\begin{center}
\includegraphics[clip,height=6cm]{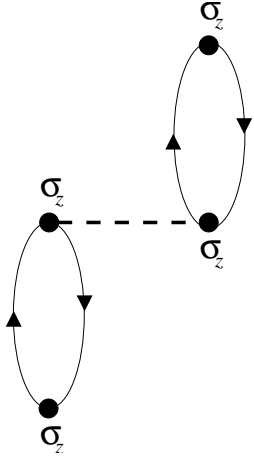}
\caption{\label{fig:piRPAdirect1}
  The first order direct term contributing to $\Pi_{zz}^{\textrm{RPA}}$.} 
\end{center}
\end{figure}

\begin{figure}
\begin{center}
\includegraphics[clip,height=4cm]{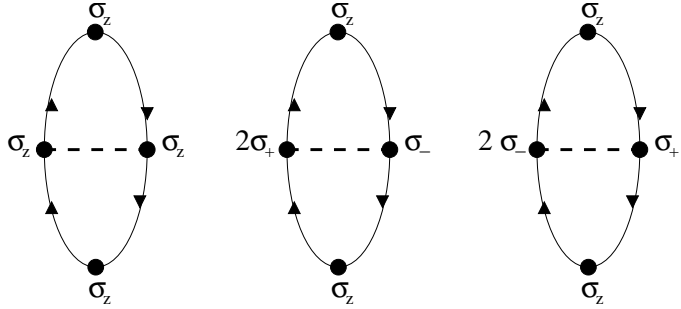}
\caption{\label{fig:piRPAexch1}
  The first order exchange terms contributing to $\Pi_{zz}^{\textrm{RPA}}$.} 
\end{center}
\end{figure}

Here we derive the exact expression for $\Pi_{zz}^{\textrm{RPA,b}}$. In
Eq.~(\ref{trick}) we made an ansatz on the form of the equation obeyed by the
latter. We have now to fix the matrix elements of the potential entering into
Eq.~(\ref{eq:Pime}). 
This can be easily done expanding Eq.~(\ref{trick}) to first order in $V_1$ and
then matching the result obtained in this way with the explicit expression
coming from the evaluation of the first order diagrams in
Figs.~(\ref{fig:piRPAdirect1}) and (\ref{fig:piRPAexch1}). 
From Eq.~(\ref{trick}) it follows immediately that
\begin{equation}
  \widehat{\Pi}=(1-\widehat{\Pi}^0\widehat{V})^{-1}\widehat{\Pi}^0.
\end{equation}
Hence the evaluation of $\widehat{\Pi}$ simply amounts to invert a $2\times 2$ 
matrix. We get:
\begin{equation}
  \widehat{\Pi} = \frac{1}{D}\left( 
    \begin{array}{cc}
      \Pi_{++}^0(1-\Pi_{--}^0 V_d) & \Pi_{++}^0 V_{od}\Pi_{--}^0 \\
      \Pi_{++}^0 V_{od}\Pi_{--}^0 & \Pi_{--}^0(1-\Pi_{++}^0 V_d)
    \end{array} \right),
\end{equation}
where
\begin{equation}
  D=1-V_d(\Pi_{++}^0+\Pi_{--}^0)+\Pi_{++}^0\Pi_{--}^0(V_d^2-V_{od}^2).
\end{equation}
The complete RPA result is thus given by
\begin{eqnarray}
  \Pi^{\textrm{RPA}} & = & \Pi_{++}+\Pi_{--}+2\Pi_{+-} \nonumber\\
  & = & \frac{1}{D}\left[\Pi_{++}^0+\Pi_{--}^0+2\Pi_{++}^0\Pi_{--}^0
    (V_{od}-V_d)\right],
\end{eqnarray}
which can be expanded to first order in $V$ obtaining
\begin{equation}
\label{match}
  \Pi_{zz}^{\textrm{(1)RPA}}=\Pi_{++}^0+\Pi_{--}^0+
    \Pi_{++}^0V_d\Pi_{++}^0+\Pi_{--}^0V_d\Pi_{--}^0+
    2\Pi_{++}^0V_{od}\Pi_{--}^0.
\end{equation}
The direct contribution to $\Pi_{zz}^{\textrm{(1)RPA}}$ comes from the diagram
in Fig.~\ref{fig:piRPAdirect1} and reads 
\begin{equation}
  \Pi_{zz}^{\textrm{(1)dir}}=\Pi_{++}^0V_1\Pi_{++}^0+\Pi_{--}^0V_1\Pi_{--}^0+
    2\Pi_{++}^0V_{od}\Pi_{--}^0.
\end{equation}
The first order exchange contribution comes from the three diagrams in
Fig.~\ref{fig:piRPAexch1} and reads 
\begin{equation}
  \Pi_{zz}^{\textrm{(1)ex}}=-\Pi_{++}^0V_1\Pi_{++}^0-\Pi_{--}^0V_1\Pi_{--}^0+
    4\Pi_{++}^0V_{od}\Pi_{--}^0.
\end{equation}
If one chooses to keep only the direct contribution to the first order result
for $\widehat{\Pi}$ (\emph{ring approximation}) then, from the matching with
Eq.~(\ref{match}), it follows that 
\begin{equation}
  V_d^{\textrm{dir}}=V_1,\; V_{od}^{\textrm{dir}}=V_1\;\Rightarrow\quad
    \widehat{V}^{\textrm{dir}}=\left( 
    \begin{array}{cc}
      V_1 & V_1 \\
      V_1 & V_1
    \end{array} \right).
\end{equation}
On the other hand, keeping only the first order exchange contribution, the
matching with Eq.~(\ref{match}) allows one to resum the \emph{ladder series}
obtaining
\begin{equation}
  V_d^{\textrm{ex}}=-V_1,\; V_{od}^{\textrm{ex}}=2V_1\;\Rightarrow\quad
    \widehat{V}^{\textrm{ex}}=\left( 
    \begin{array}{cc}
      -V_1 & 2V_1 \\
      2V_1 & -V_1
    \end{array} \right).
\end{equation}  
Finally, from the matching of Eq.~(\ref{match}) with all the four diagrams in
Figs.~(\ref{fig:piRPAdirect1}) and (\ref{fig:piRPAexch1}), one gets the
\emph{full RPA result} (direct + exchange):
\begin{equation}
  V_d^{\textrm{RPA}}=0,\; V_{od}^{\textrm{RPA}}=3V_1\;\Rightarrow\quad
    \widehat{V}^{\textrm{RPA}}=\left( 
    \begin{array}{cc}
      0 & 3V_1 \\
      3V_1 & 0
    \end{array} \right).
\end{equation} 

\section{The HF polarization propagator in a broken vacuum}
\label{app:im_re_pi_+-}

In this appendix we display how the calculations leading to 
$\Pi_{\mp\pm}^{\textrm{HF}}$ (which enter into Eq.~(\ref{resummed}) for 
$\Pi_{\mp\pm}^{\textrm{RPA}}$) can be performed analytically in a quite
straightforward way following a procedure first introduced in
Ref.~\cite{DeP98}.
Starting from the expression for $\Pi_{-+}^{\textrm{HF}}$ given in 
Eq.~(\ref{eq:Pi+}), one introduces the \emph{particle-hole propagator} 
in the HF approximation, defined as:
\begin{equation}
  G_\textrm{ph}^{\textrm{HF}}(\vec{k},\vec{q},\omega)
    =\frac{\theta(|\vec{k}+\vec{q}|-k_F^-)\theta(k_F^+ -k)}
    {\omega+\omega_{\vec{k}}^{+}-\omega_{\vec{k}+\vec{q}}^{-}
    +i\eta}-\frac{\theta(k_F^{-}-|\vec{k}+\vec{q}|)\theta(k-k_F^{+})}
    {\omega+\omega_{\vec{k}}^{+}-\omega_{\vec{k}+\vec{q}}^{-}-i\eta}.
\end{equation}
The above expression can be manipulated in the following way:
\begin{eqnarray}
  G_\textrm{ph}^{\textrm{HF}}(\vec{k},\vec{q},\omega) & = & 
    \frac{\theta(|\vec{k}+\vec{q}|-k_F^-)\theta(k_F^+ -k)}
    {\omega+\omega_{\vec{k}}^{+}-\omega_{\vec{k}+\vec{q}}^{-}
    +i\eta}+\frac{\theta(k_F^{-}-|\vec{k}+\vec{q}|)\theta(k-k_F^{+})}
    {-\omega-\omega_{\vec{k}}^{+}+\omega_{\vec{k}+\vec{q}}^{-}+i\eta}
    \nonumber\\
  && +\frac{\theta(k_F^- -|\vec{k}+\vec{q}|)\theta(k_F^+ -k)}
    {\omega+\omega_{\vec{k}}^{+}-\omega_{\vec{k}+\vec{q}}^{-}
    +i\eta_\omega}+\frac{\theta(k_F^- -|\vec{k}+\vec{q}|)\theta(k_F^+ -k)}
    {-\omega-\omega_{\vec{k}}^{+}+\omega_{\vec{k}+\vec{q}}^{-}-i\eta_\omega}
    \nonumber\\
  &=& \frac{\theta(k_F^+ -k)-\theta(k_F^- -|\vec{k}+\vec{q}|)}
    {\omega-\omega_{\vec{k}+\vec{q}}^- +\omega_{\vec{k}}^+ +i\eta_\omega},
\end{eqnarray}
being $\eta_\omega=\textrm{sign}(\omega)\eta$.
Hence, expressing $\Pi_{-+}^{\textrm{HF}}$ in terms of the particle-hole 
propagator
\begin{equation}
  \Pi^{\textrm{HF}}_{-+}(q,\omega)=\int\frac{{\mbox{d}}\vec{k}}{(2\pi)^3}
    G_\textrm{ph}^{\textrm{HF}}(\vec{k},\vec{q},\omega)=
    \int\frac{{\mbox{d}}\vec{k}}{(2\pi)^3}\frac{\theta(k_F^+ -k)-\theta
    (k_F^- -|\vec{k}+\vec{q}|)}{\omega-\omega_{\vec{k}+\vec{q}}^- 
    +\omega_{\vec{k}}^+ +i\eta_\omega},
\end{equation}  
and performing the change of variable $\vec{k}+\vec{q}\to\vec{k}$ 
within the integral for the second contribution, one gets:
\begin{equation}
  \Pi^{\textrm{HF}}_{-+}(q,\omega)=\int\frac{{\mbox{d}}\vec{k}}{(2\pi)^3}
    \left[\frac{\theta(k_F^+ -k)}{\omega-\omega_{\vec{k}+\vec{q}}^- 
    +\omega_{\vec{k}}^+ +i\eta_\omega}-\frac{\theta(k_F^- -k)}
    {\omega+\omega_{\vec{k}+\vec{q}}^+ -\omega_{\vec{k}}^- +i\eta_\omega}
    \right].
\end{equation}
The energy denominators in the equation above can be written as
\begin{subequations}
\begin{eqnarray}
  \omega-\omega_{\vec{k}+\vec{q}}^- +\omega_{\vec{k}}^+ &=& 
    \omega-\frac{q^2}{2m}-\frac{\vec{q}\cdot\vec{k}}{m}-\Delta\omega \\
  \omega+\omega_{\vec{k}+\vec{q}}^+ -\omega_{\vec{k}}^- &=&
    \omega+\frac{q^2}{2m}+\frac{\vec{q}\cdot\vec{k}}{m}-\Delta\omega,
\end{eqnarray}
\end{subequations}
where
\begin{equation}
  \Delta\omega=-\frac{V_1}{2\pi^2}\left[(k_F^+)^3-(k_F^-)^3\right].
\end{equation}
Note that at equilibrium $V_1$ is related to $k_F^{\pm}$ through
Eq.~(\ref{eq:zerorange}) and one gets to the expression (\ref{eq:Deltaomega})
of Section~\ref{sec:xaxis}.

It is particularly convenient to introduce the \emph{scaling variables}:
\begin{subequations}
\begin{eqnarray}
  \psi_{\pm}&=&\frac{1}{k_F^\pm}\left(\frac{m\omega}{q}-\frac{q}{2}\right) \\
  \Delta\psi_{\pm}&=&\frac{m\Delta\omega}{k_F^\pm q}.
\end{eqnarray}
\end{subequations}
In terms of the above variables we can write:
\begin{subequations}
\begin{eqnarray}
  \omega-\omega_{\vec{k}+\vec{q}}^- +\omega_{\vec{k}}^+ &=& \frac{k_F^- q}{m}
    \left(\psi_- - \Delta\psi_- -\frac{\hat{q}\cdot\vec{k}}{k_F^-}\right) \\
  \omega+\omega_{\vec{k}+\vec{q}}^+ -\omega_{\vec{k}}^- &=& \frac{k_F^+ q}{m}
    \left(\psi_+ - \Delta\psi_+
    +\frac{q}{k_F^+}+\frac{\hat{q}\cdot\vec{k}}{k_F^+}\right). 
\end{eqnarray}
\end{subequations}
Hence $\Pi_{-+}^{\textrm{HF}}$ can be expressed analytically as follows:
\begin{equation}
  \Pi_{-+}^{\textrm{HF}}(q,\omega)=\frac{m}{q}\frac{1}{(2\pi)^2}
    \left[(k_F^-)^2\mathcal{Q}^{(0)}(\psi_- - \Delta\psi_-)-(k_F^+)^2
    \mathcal{Q}^{(0)}(\psi_+ - \Delta\psi_+ +\bar{q}_+)\right],
\end{equation} 
being $\bar{q}_\pm=q/k_F^\pm$.
Analogously, for $\Pi_{+-}^{\textrm{HF}}$ one gets:
\begin{equation}
  \Pi_{+-}^{\textrm{HF}}(q,\omega)=\frac{m}{q}\frac{1}{(2\pi)^2}
    \left[(k_F^+)^2\mathcal{Q}^{(0)}(\psi_+ + \Delta\psi_+)-(k_F^-)^2
    \mathcal{Q}^{(0)}(\psi_- + \Delta\psi_- +\bar{q}_-)\right].
\end{equation}
The expressions reported above allow us to write
$\textrm{Re}\Pi_{\mp}^{\textrm{HF}}$ ($\textrm{Re}\Pi_{\pm}^{\textrm{HF}}$) and
$\textrm{Im}\Pi_{\mp}^{\textrm{HF}}$ ($\textrm{Im}\Pi_{\pm}^{\textrm{HF}}$) in
terms of Legendre polynomials and functions of second kind, $P_n$ and $Q_n$,
respectively. Indeed it is easy to show that
\begin{subequations}
\begin{eqnarray}
  \textrm{Re}\mathcal{Q}^{(0)}(\psi)&=&\frac{2}{3}[Q_0(\psi)-Q_2(\psi)] \\
  \textrm{Im}\mathcal{Q}^{(0)}(\psi)&=&-\textrm{sign}(\omega)\theta(1-\psi^2)
    \frac{\pi}{3}[P_0(\psi)-P_2(\psi)].
\end{eqnarray}
\end{subequations}

\end{document}